\documentclass[traditabstract]{aa}
\usepackage{epsfig}    
\usepackage{lscape}
\usepackage{natbib}
\bibpunct{(}{)}{;}{a}{}{,}    %% natbib format like A&A and ApJ
\usepackage{graphics}
\usepackage{graphicx,graphics}
\usepackage{color} 
\usepackage{times}
\usepackage{amsmath}
\usepackage{amssymb}
\usepackage[colorlinks=true,linkcolor=blue,citecolor=blue,filecolor=blue,urlcolor=blue]{hyperref}

\begin{document}

\title{Central oxygen abundances in the spiral galaxies of the MaNGA survey: Galaxies with central starbursts} 
      
\author{
         L.~S.~Pilyugin\inst{\ref{ITPA},\ref{MAO}}           \and 
         G.~Tautvai\v{s}ien\.{e}\inst{\ref{ITPA}}    
}
\institute{Institute of Theoretical Physics and Astronomy, Vilnius University, Sauletekio av. 3, 10257, Vilnius, Lithuania \label{ITPA} 
\and
  Main Astronomical Observatory, National Academy of Sciences of Ukraine, 27 Akademika Zabolotnoho St, 03680, Kiev, Ukraine \label{MAO}
}

\abstract{
  We examine whether there are deviations of the local central oxygen abundances in spiral galaxies from the general metallicity gradients.
  We compare the values of the central
  intersect oxygen abundances estimated from the metallicity gradient based on the integral field unit (IFU) spectroscopy from the Mapping Nearby Galaxies
  at the Apache Point Observatory (MaNGA) survey and the local central oxygen abundances obtained from the single-fibre observations from the Sloan Digital Sky Survey (SDSS).
  Special attention is placed on galaxies with recent and currently ongoing central starbursts (cSB galaxies). We selected a sample of 30 cSB galaxies from our total sample of 381
  MaNGA galaxies, using the decrease in the  D$_{n}$4000 index (a stellar age indicator) in the circumnuclear region as the selection criterion. We found that the local
  central oxygen abundances follow  the general metallicity gradients in the galaxies well and agree with the central intersect abundances within uncertainties of the central
  abundances determinations. Starbursts in the centres of cSB galaxies do not produce noticeable oxygen enrichments. The central
  starbursts imply that an appreciable  amount of gas is present at the centres of cSB galaxies. The gas at the centre of galaxy  can serve not only as a raw material for the
  star formation, but also as a fuel for the activity of the galactic nucleus (AGN). We found that the AGN is the main source of the ionising radiation at the centres of six 
  cSB galaxies in our sample.
}


\titlerunning{Central oxygen abundances in spiral MaNGA galaxies}
\authorrunning{Pilyugin and Tautvai{\v s}ien\.{e}}

\maketitle

\section{Introduction}
%=====================

It has been known for some time that the discs of spiral galaxies exhibit negative radial abundance gradients in the sense that the abundance is higher at the centre and decreases with
the galactocentric distance \citep{Searle1971,Smith1975}.  
The characteristic abundances are used to compare the abundance properties among a sample of spiral galaxies.
The oxygen abundance at the centre of a galaxy is the most frequently used choice for the characteristic abundance
\citep[][among many others]{VilaCostas1992,Tremonti2004,Pilyugin2004,Pilyugin2014,Pilyugin2019,Curti2020}.
Alternatively, the value of oxygen abundance at the effective  (half-light) radius as a characteristic metallicity of the galaxy is also used in many investigations  
\citep[e.g.][]{Garnett1987,Garnett2002,Sanchez2013,Sanchez2019,Alvarez-Hurtado2022,Sanchez-Menguiano2024}.
The value of oxygen abundance at other radiuses can be also used as a characteristic metallicity of a galaxy. For instance,  
\citet{Zaritsky1994} defined the characteristic oxygen abundance as the abundance at 0.4~$R_{25}$, where $R_{25}$ is the optical (or isophotal) radius of
the galaxy.  This galactocentric distance is close to the effective radius of the galaxy.

Integral field unit (IFU) spectroscopy for a large sample of spiral galaxies was obtained within a framework of the Mapping Nearby Galaxies at Apache Point Observatory
survey (MaNGA, \citet{Bundy2015}). Those measurements provide a possibility to determine a distribution of oxygen abundances across the galactic discs and establish
radial oxygen abundance distributions.
In our previous study \citep{Pilyugin2024}, we considered two sequences of well measured spiral galaxies from the MaNGA survey with different shapes of the radial oxygen abundance
distributions. The gradients in those galaxies were approximated by a single or broken linear relation. The binned oxygen abundances (the median values of the abundances in bins of
0.05~dex in $R/R_{25}$) were used instead of the oxygen abundances in individual spaxels in order to minimise the influence of the spaxels
with unreliable abundances in determining the radial abundance distribution. The majority of the binned oxygen abundances are close to the O/H -- $R$ relation, with the mean deviation in the O/H from the relation being less than $\sim$0.01~dex. However, appreciable deviations from the  O/H -- $R$ relation can take place in the spaxels near the 
galactic centres. These deviations can be attributed to uncertainties in the oxygen abundance
determinations in those  spaxels \citep{Pilyugin2024}. Therefore, those spaxels are rejected in the determination of the radial abundance gradient, and the central oxygen abundance
in a galaxy is estimated as the intersect value of the determined radial abundance distribution.
\citet{Belfiore2017} noted that the observed metallicities in spaxels near the galactic centre  can be affected by the ‘beam-smearing’ effect of a  point spread function (PSF).
The point spread function of the MaNGA measurements is estimated to have a full width at the half maximum (FWHM) of 2.5~arcsec or five pixels \citep{Bundy2015,Belfiore2017}. For low $R_{25}$/PSF ratios
and high galaxy inclinations, the error in the oxygen abundance at the centre can be as high as $\sim$0.04~dex. 

On the other hand, it cannot be excluded that the deviation of the central oxygen abundance from the general radial gradient can be real under some conditions. Indeed, some giant spiral
galaxies undergo star formation bursts in their centres or have undergone such events in the recent past. It is important to investigate whether such galaxies have had their central oxygen abundances altered.  
Values of the D$_{n}$(4000) index (which is an indicator of the stellar age) increase with the decrease in the  radius in the central regions of majority of giant spiral galaxies. In a number
of giant spiral galaxies, the  values of the D$_{n}$(4000) index at the central region increase with decreasing galactocentric distance up to some radius, and then start to decrease towards the centre.
This suggests an appreciable recent star formation in centres of those galaxies. We refer to those galaxies as central starburst galaxies (cSB galaxies). If the star burst alters
the (local) oxygen abundance at the centre of spiral galaxy,
then the intersect oxygen abundance does not correspond to the real central abundance in the cSB galaxy. Unfortunately,  the uncertainties in the oxygen abundances in
spaxels near the centre estimated on the base of the MaNGA observations are an obstacle to distinguishing whether the deviation of the observed central abundance from the general gradient is real or false.

A single-fibre spectroscopy of centres for a large number of galaxies (including the MaNGA galaxies) was carried out within the  Sloan Digital Sky Survey (SDSS, \citet{York2000}).
A comparison between the intersect values of the central abundances based on the MaNGA measurements and oxygen abundances estimated from the SDSS measurements offers an
opportunity to reveal  the deviation in the oxygen abundance at the centre of spiral galaxy from the general radial metallicity gradient, if such a deviaton does indeed exist.

A significant star formation rate at the galactic centre implies that  an appreciable  amount of gas is present at (or flows on) the centre of galaxy.   
It is believed \citep[e.g.][]{Heckman2014} that the nuclei of most, and quite possibly all, massive galaxies host a central supermassive black hole (SMBH).
The gas at the centre of galaxy  can serve not only as the raw material for the star formation but also as the fuel to the active galactic nucleus (AGN). 
We can expect that both main excitation mechanisms (i.e. starburst or AGN) can be found at the centres of giant spiral cSB galaxies.  
The classification of spaxel spectra using the Baldwin-Phillips-Terlevich diagnostic diagram (BPT diagram, \citet{Baldwin1981}) and examination of distributions of
spaxels with different BPT types of the radiation in the circumnuclear regions can reveal an interplay between the star formation and AGN in those galaxies.
{The use of the equivalent width of the emission  H$\alpha$ line versus  velocity dispersion diagram (WHaD diagram, \citep{Sanchez2024}) as an alternative diagnostic
  of different ionising sources  can demonstrate an interplay between the star formation and AGN in those galaxies.  This can also clarify the origin of the deviations of
  the oxygen abundances in spaxels near the centre from the general gradient in some MaNGA galaxies.}

In this work,  we examine whether there are deviations of the local central oxygen abundances in spiral galaxies from the general metallicity gradients.
We verify the validity of the central intersect oxygen abundances determined from the MaNGA observations by comparing
them with the central oxygen abundances estimated from the fibre SDSS spectra. Special attention is paid to the cSB galaxies.  {We examine sources
  of the ionising radiation  (starburst or AGN) at the centres of cSB galaxies in two ways: (1) by constructing  the distributions of spaxels with different BPT types
  of the radiation in the circumnuclear regions and (2) by using an alternative diagnostic (WHaD diagram) of different ionising sources.} The paper is organised in the following way:
the data are described in Section 2, Section 3 presents a discussion, and Section 4 provides a summary.

\section{Data}
%=====================

\subsection{Sample} 
%=====================

Our investigation is based on galaxies from the MaNGA survey \citep{Bundy2015}. Using the publicly available
spectroscopy (Data Release 17, \citealt{Abdurrouf2022}),  we determined maps and radial distributions of properties in 381 MaNGA galaxies. For the current study, we selected
a sample of galaxies using the  criteria described below.

We selected galaxies whose curves for the isovelocities in the measured gas velocity fields correspond to a set of parabola-like curves (i.e. an hourglass-like appearance
of the rotation disc). This condition offers the possibility of selecting a sample of disc galaxies and to determine the orientation of the galaxy in space (the pixel coordinates
of the galactic rotation centre, position angle of the major kinematic axis, and inclination angle of a galaxy) and the rotation curve. These geometrical parameters
are necessary in determination of the galactocentric distances of spaxels. This criterion also allows us to reject strongly interacting and merging galaxies. 

Galaxies with an inclination angle larger than $\sim$70$\degr$ were rejected because a fit of the H$\alpha$ velocity field in galaxies with a high ratio of
the major to minor axis can produce unrealistic values of the inclination angle \citep{Epinat2008}; as a result, the estimated galactocentric distances of the spaxels
located far from the major axis could then involve large uncertainties. In addition, the interpretation of the abundance in an individual spaxel may be questioned because
the spaxel spectra involve radiations that originated at different galactocentric distances.  At the same time, galaxies with a small inclination (face-on galaxies) are
included in our sample. The inclination angle determined for the face-on galaxy may involve a significant uncertainty. The uncertainty in the inclination angle of the face-on
galaxy results in a small uncertainty in the galactocentric distances of spaxels, but results in a significant uncertainty in the obtained rotation velocity. Therefore,
the rotation velocities are not considered in the current study. Instead, the stellar mass of the galaxy and its size (optical radius) are used as global parameters. 

The MaNGA galaxies mapped with 91 and 127-fibre IFU, covering  27$\farcs$5 and 32$\farcs$5 on the sky (with a large number of spaxels over the galaxy image) are taken into
consideration. Only the galaxies where the spaxels with measured emission lines are well distributed across galactic discs, covering more than $\sim 0.75~R_{25}$, are considered.
This condition provides a possibility to estimate more or less reliable values of the kinematic angles, the optical radius, and the radial distributions of different characteristics.

The emission line parameters of the spaxel spectra for our sample of galaxies were taken from the MaNGA Data Analysis Pipeline (DAP) measurements. We determined the characteristics
of out sample of galaxies using the last version of
DAP measurements: `manga-n-n-MAPS-SPX-MILESHC-MASTARSSP.fits.gz'\footnote{https://data.sdss.org/sas/dr17/manga/spectro/analysis/v3\_1\_1/3.1.0/SPX-MILESHC-MASTARSSP/} for the spectral data 
and the data reduction pipeline (DRP) measurements: `manga-n-n-LOGCUBE.fits.gz'\footnote{https://dr17.sdss.org/sas/dr17/manga/spectro/redux/v3\_1\_1/} for the photometric data.
Determinations of the geometrical parameters of galaxies (i.e. coordinates  of the centre,  position angle of the major axis,  inclination angle, and isophotal radius),
construction of the maps, and the radial distributions of different characteristics across galaxies were carried out in the same manner described in our previous papers
\citep{Pilyugin2018,Pilyugin2019,Pilyugin2021,Pilyugin2024}.

\subsection{Sample of galaxies with central starbursts (cSB galaxies)}
%=====================

%===============    Fig  No 1          Dn4000 distribution examples 
\begin{figure}
\resizebox{1.00\hsize}{!}{\includegraphics[angle=000]{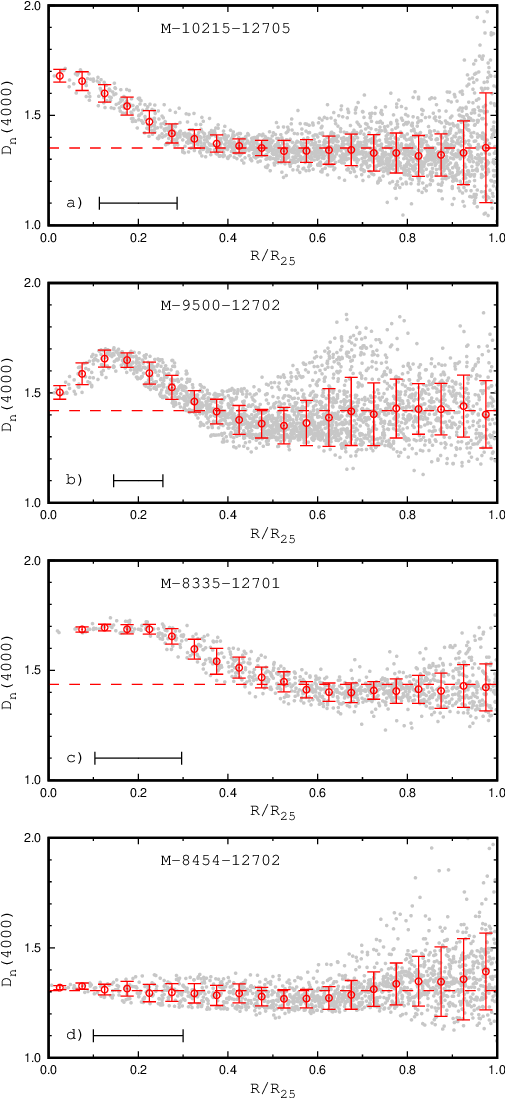}}
\caption{ Examples of different behaviour of the D$_{n}$(4000) indices along the galactic radius.
  {\em Panel} {a:} M-10215-12705, an example of galaxy with the increasing D$_{n}$(4000) towards the centre at the central region. 
  {\em Panels} {b,c:} M-9500-12702 (panel a) and M-8335-12701 (panel b), examples of galaxies where the D$_{n}$(4000) at the central region increases up to some radius
  and then decreases towards the centre. 
  {\em Panel} {d:} M-8454-12702, an example of the galaxy with a small variation in the values of D$_{n}$(4000) indices along the radius.
  In each panel, the grey points denote D$_{n}$(4000) indices in the individual spaxels, the red circles mark the median values of
  the D$_{n}$(4000) in bins of 0.05 in the fractional radius $R/R_{25}$, the bars show a scatter in D$_{n}$(4000) about the median value in the bins, and the red dashed line shows
  the median value of the  D$_{n}$(4000) indices in all spaxels within the optical radius. {The horizontal bar indicates the point spread function in the MaNGA measurements.}
}
\label{figure:age-distribution-examples}
\end{figure}

The D$_{n}$(4000) index is an indicator of the stellar age. The decrease in D$_{n}$(4000) at the galactic centre offers evidence that an appreciable star formation occur at the centre
of the galaxy recently or at the recent past. Thus, the decrease in D$_{n}$(4000) at the centre of the galaxy can be used as a criterion to select galaxies with the central
starbursts (cSB galaxy).  There is also another possibility to determine the (luminosity-weighed) ages of stellar populations in spaxels of the MaNGA galaxies; for instance, as applied by
\citet{Sanchez2022},  namely, by ﬁfitting the spaxel spectrum with a combination of simple stellar population spectra. \citet{Pilyugin2024} compared the stellar ages based on the spectral
index D$_{n}$(4000)  and luminosity-weighed ages of the stellar populations in spaxels of the MaNGA galaxies and showed that the behaviours of the values of
both stellar ages along the radius is similar. Thus, in this work, we used the radial distribution of the D$_{n}$(4000) indices to select a sample of cSB galaxies.

In Fig.~\ref{figure:age-distribution-examples}, we show examples of three different types of the D$_{n}$(4000) index behaviours along galactic radius.
The position of the spaxel in the disc is specified by the fractional radius, $R_{g}$, normalised to the isophotal (or optical) radius  $R_{g}$ = $R/R_{25}$.
Panel (a) in Fig.~\ref{figure:age-distribution-examples} shows the D$_{n}$(4000) index as a function of radius for the MaNGA galaxy M-10215-12705, which is
a Sbc galaxy with a stellar mass of log($M_{\star}/M_{\sun}$) = 11.087 and a physical optical radius of $R_{25}$ = 22.83~kpc.
Then, M-10215-12705 is an example of galaxy with the increase in D$_{n}$(4000) towards the centre in the central region.  This behaviour of  D$_{n}$(4000) along the radius
is typical for giant spiral galaxies.

Panel (b) in Fig.~\ref{figure:age-distribution-examples} shows D$_{n}$(4000) as a function of the radius for the MaNGA galaxy M-9500-12702, which is a SBab galaxy with 
the stellar mass of log($M_{\star}/M_{\sun}$) = 10.744 and the physical optical radius of $R_{25}$ = 13.68~kpc.
M-9500-12702 is an example of galaxy  where D$_{n}$(4000) at the central region increases with decrease in the radius up to a certain value,  and then decreases towards the centre.
Panel (c) in Fig.~\ref{figure:age-distribution-examples} shows another example of galaxy  where D$_{n}$(4000) at the central region increases with decrease in the radius up to a certain value, and then decreases towards the centre. The MaNGA galaxy M-8335-12701  is a galaxy of
the stellar mass of log($M_{\star}/M_{\sun}$) = 10.973 and of the physical optical radius of $R_{25}$ = 17.92~kpc.
Panel (d) in Fig.~\ref{figure:age-distribution-examples} shows D$_{n}$(4000) as a function of radius for the MaNGA galaxy M-8454-12702, which is a SBbc galaxy with 
the stellar mass of log($M_{\star}/M_{\sun}$) = 11.136 and of the physical optical radius of $R_{25}$ = 20.67~kpc. Then,
M-8454-12702 is an example of galaxy  with a small variation in the D$_{n}$(4000) indices along the radius.

It should be noted that the scatter in the D$_{n}$(4000) indices in individual spaxels at a given galactocentric distance can significantly increase with the radius  and the binned 
values can increase along the radius in the outer parts in many galaxies.  We can assume that measurements of D$_{n}$(4000) in individual spaxels in the outer parts of galaxies
can involve large uncertainties. The increase in the binned D$_{n}$(4000)  with the radius (at least a major part of this increase) is not real and ought to be attributed to the
uncertainties in D$_{n}$(4000) measurements in individual spaxels. 

We selected a sample of 30 cSB galaxies in which  the D$_{n}$(4000) indices at central regions increase with the decrease in the radius up to some radius, then decrease towards the centre
(radial D$_{n}$(4000) distributions shown in the panels (b) and (c) of  Fig.~\ref{figure:age-distribution-examples}). 
The radial distributions of the D$_{n}$(4000) indices,  the suface brightness in the H$\alpha$ emission line, the oxygen abundances and maps of the distribution of the radiation of
different BPT types across the image of galaxy for our sample of the cSB galaxies are shown in Figs.~\ref{figure:app-fig1}-~\ref{figure:app-fig5} in the appendix. 

Table~\ref{table:general} lists the general characteristics of each cSB galaxy of our sample. 
  The first column gives the MaNGA name for each galaxy.
  The distance is reported in Column 2. The distances to the galaxies were adopted from the NASA/IPAC Extragalactic Database ({\sc ned})\footnote{The NASA/IPAC Extragalactic Database
({\sc ned}) is operated by the Jet Populsion Laboratory, California Institute of Technology, under contract with the National Aeronautics and Space Administration.
{\tt http://ned.ipac.caltech.edu/}}.  The {\sc ned} distances use the flow corrections for Virgo, the Great Attractor, and Shapley Supercluster infall (adopting a
cosmological model with $H_{0} = 73$~km/s/Mpc, $\Omega_{m} = 0.27$, and $\Omega_{\Lambda} = 0.73$). The reported errors in distances are less than 10\%.  
  The stellar mass is given in Column 3. We have chosen the spectroscopic $M_{sp}$ masses of the SDSS
and Baryon Oscillation Spectroscopic Survey
in SDSS-III \citep[BOSS, see][]{Dawson2013}.  The spectroscopic masses are taken from the
table {\sc stellarMassPCAWiscBC03}, and were determined using the Wisconsin method \citep{Chen2012} with the stellar population synthesis models from \citet{Bruzual2003}.
The reported errors in the values of the stellar mass are usually within 0.15 -- 0.2~dex. 
{
The isophotal radius  $R_{25}$ in arcsecond, determined here, is listed in Column 4.
The isophotal radius in kiloparsecs, obtained from the isophotal radius in arcsecond (Column 4) and the adopted distance (Column 2), is listed in Column 5.
}
  The central intersect oxygen abundance 12+log(O/H)$_{0,{\rm intersect}}$ estimated from the metallicity gradient determined on the base of the IFU spectroscopy from the MaNGA
is reported in Column 6.
  The local central oxygen abundance 12+log(O/H)$_{\rm SDSS}$ obtained from the single-fibre observations from the SDSS is given in Column 7.
  The configuration of the distribution of the BPT types of the radiation at the centre (SF -- H\,{\sc ii} region-like spectra at the centre, INT -- intermediate type spectra,
AGN -- AGN-like spectra, SF+INT -- the inner region of the   H\,{\sc ii}-region-like radiation is surrounded by a ring of radiation of the intermediate type) is reported
in Column 8.
The presence of a bar at the centre of galaxy according to the HyperLeda\footnote{http://leda.univ-lyon1.fr/} database \citep{Makarov2014} is indicated in Column 9.
The environment (isolated galaxy or a group member) according to \citet{Tempel2018} is reported in Column 10.

%++++++++++++++++++ Table 1    General characteristics 
%\setcounter{table}{0}
\begin{table*}
\caption[]{\label{table:general}
    General characteristics of our sample of the MaNGA galaxies with central starbursts (cSB galaxies)
}
\begin{center}
\begin{tabular}{cccccccccc} \hline \hline
%\multicolumn{3}{|c|}{scatter in 12+logO/H)$_{R_{x}}$}      \\           
MaNGA name                &
distance                  &
log\,$M_{\star}$            &
$R_{25}$                   &
$R_{25}$                   &
12+log(O/H)$_{0.{\rm intersect}}$  &
12+log(O/H)$_{\rm SDSS}$       &
BPT$^{1}$                  &
Bar                       &
environment               \\
                       &
Mpc                    &
$M_{\sun}$              &
arcsecond              &
Kpc                    &
                       &
                       &
                       &
                       &
                      \\
\hline
%                dMpc       Msp      R25      R25       OHinter    OHsdss   BPT     Bar     neighbour 
  7960 12705 &  137.9 &   10.552 &  19.3  &  12.90 &    8.663 &    8.678 & SF     &       & multu     \\
  8132 12705 &  250.9 &   10.945 &  10.4  &  12.65 &    8.667 &          & SF+INT &  yes  & multu     \\
  8137 09102 &  139.2 &   10.670 &  22.6  &  15.29 &    8.660 &    8.657 & SF+INT &  yes  & isolated  \\
  8335 12701 &  286.6 &   10.973 &  12.9  &  17.92 &    8.627 &    8.660 & SF+INT &       & isolated  \\
  8453 12701 &  115.6 &   10.325 &  22.3  &  12.50 &    8.639 &    8.661 & SF+INT &       & multu     \\
  8482 12705 &  190.1 &   11.142 &  21.8  &  20.04 &    8.637 &    8.655 & SF+INT &       & multu     \\
  8549 12702 &  197.7 &   11.235 &  19.0  &  18.21 &    8.627 &          & AGN    &  yes  & multu     \\
  8569 12701 &  250.9 &   10.826 &  12.8  &  15.63 &    8.674 &    8.670 & SF+INT &       &           \\
  8724 12701 &  142.3 &   10.749 &  24.4  &  16.83 &    8.671 &          & SF+INT &  yes  & isolated  \\
  8988 12704 &  264.1 &   10.815 &  11.4  &  14.53 &    8.660 &          & INT    &  yes  & multu     \\
  8990 12703 &  249.6 &   10.674 &  13.4  &  16.28 &    8.636 &    8.646 & SF     &       & multu     \\
  9026 12703 &  324.4 &   11.370 &  11.6  &  18.32 &    8.627 &          & SF     &       & multu     \\
  9028 12705 &  276.8 &   10.917 &  10.8  &  14.56 &    8.648 &          & AGN    &       & isolated  \\
  9031 12705 &  270.6 &   10.964 &  10.7  &  13.97 &    8.669 &          & AGN    &       & multu     \\
  9038 12703 &  317.9 &   11.294 &  15.4  &  23.73 &    8.666 &    8.650 & SF+INT &  yes  & isolated  \\
  9191 09102 &  134.3 &   10.886 &  23.1  &  15.04 &    8.632 &    8.628 & SF     &       &           \\
  9484 12703 &  142.5 &   11.048 &  21.4  &  14.75 &    8.624 &          & AGN    &  yes  & multu     \\
  9500 12702 &  125.7 &   10.744 &  22.4  &  13.68 &    8.638 &    8.629 & SF+INT &  yes  & isolated  \\
  9882 12701 &  132.3 &   10.452 &  15.0  &   9.62 &    8.629 &    8.638 & SF     &       & multu     \\
 10507 12705 &  248.1 &   10.930 &  12.6  &  15.09 &    8.696 &          & SF+INT &  yes  & multu     \\
 10845 12704 &  215.0 &   10.455 &  12.8  &  13.39 &    8.620 &    8.626 & SF     &       & isolated  \\
 11757 12703 &  298.3 &   11.419 &  18.1  &  26.18 &    8.616 &          & AGN    &       & multu     \\
 11830 09102 &  252.3 &   11.553 &  20.9  &  25.56 &    8.675 &          & AGN    &       & isolated  \\
 11831 12703 &  136.1 &   10.722 &  21.8  &  14.38 &    8.605 &    8.589 & SF+INT &  yes  & multu     \\
 11872 12702 &  271.1 &   11.006 &  11.8  &  15.51 &    8.697 &    8.669 & SF+INT &  yes  & isolated  \\
 11872 12704 &  224.7 &   10.556 &  13.9  &  15.14 &    8.640 &    8.664 & SF     &       & multu     \\
 12074 12703 &  184.0 &   10.840 &  22.2  &  19.76 &    8.642 &    8.665 & SF+INT &  yes  &           \\
 12074 12704 &  192.3 &   11.160 &  17.5  &  16.36 &    8.672 &    8.648 & SF+INT &  yes  &           \\
 12488 12704 &  261.0 &   10.917 &  16.9  &  21.38 &    8.655 &          & INT    &       & isolated  \\
 12624 12703 &  281.9 &   10.970 &  14.7  &  20.09 &    8.615 &    8.653 & SF+INT &  yes  & multu     \\
                    \hline
\end{tabular}\\
\end{center}
Notes: BPT types of radiation at the centre: SF -- H\,{\sc ii} region-like spectra at the centre, INT -- intermediate type spectra, AGN -- AGN-like spectra,
SF+INT -- the inner region of the   H\,{\sc ii}-region-like radiation is surrounded by a ring of radiation of the intermediate type. 
\end{table*}

  The enhancement of specific star formation in central regions of galaxies can be attributed to the galaxy-galaxy interactions  \citep[e.g.][]{BarreraBallesteros2015,Guo2016,Thorpl2019}. 
We examined whether the central starbursts in galaxies of our sample are linked to their present-day environment or to the presence of the bar. Overall, 26 cSB galaxies of our sample are in
a catalogue of galaxy groups and clusters \citep{Tempel2018}; furthermore,16 cSB galaxies are members of galaxy groups, and 10 galaxies are classified as isolated galaxies.
The bars present at the centres of 14 cSB galaxies (in 7 isolated galaxies and in 7 galaxies that are members of the groups) according to the morphological classification from the HyperLeda.
Thus, there is no  distinctive link of the central starburst in a galaxy to its present-day environment or to the presence of the bar. This suggests that either there is
another reason responsible for the central starburst; alternatively, the central starbursts are not uniform in their origin and may instead originate from different pathways.

\section{Discussion}
%=====================

\subsection{Radiation of different BPT types in the circumnuclear regions of cSB galaxies} 
%=====================

%===============    Fig  No 2          BPT example 
\begin{figure}
\resizebox{1.00\hsize}{!}{\includegraphics[angle=000]{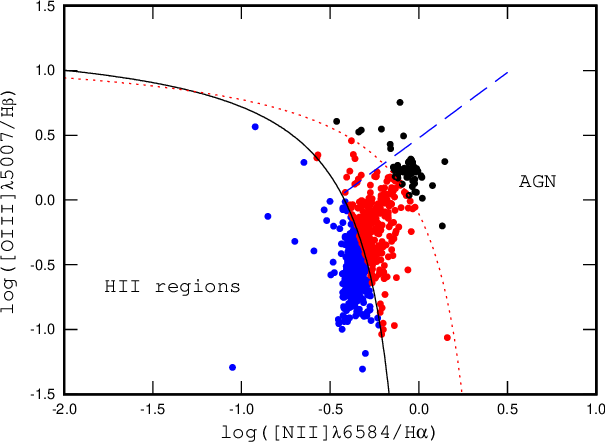}}
\caption{
  Example of BPT diagrams for the individual spaxels in a galaxy.
The individual spaxels with the H\,{\sc ii}-region-like spectra are denoted by the blue symbols,
the spaxels with AGN-like spectra are shown  by the dark symbols, and the red symbols are the spaxels with
intermediate spectra. The solid and short-dashed curves mark the demarcation line between AGNs and H\,{\sc ii}
regions defined by \citet{Kauffmann2003} and \citet{Kewley2001}, respectively. The long-dashed line is the
dividing line between Seyfert galaxies and LINERs defined by \citet{CidFernandes2010}.
}
\label{figure:bpt-example}
\end{figure}

%===============    Fig  No 3          Four configurations 
\begin{figure}
\resizebox{1.00\hsize}{!}{\includegraphics[angle=000]{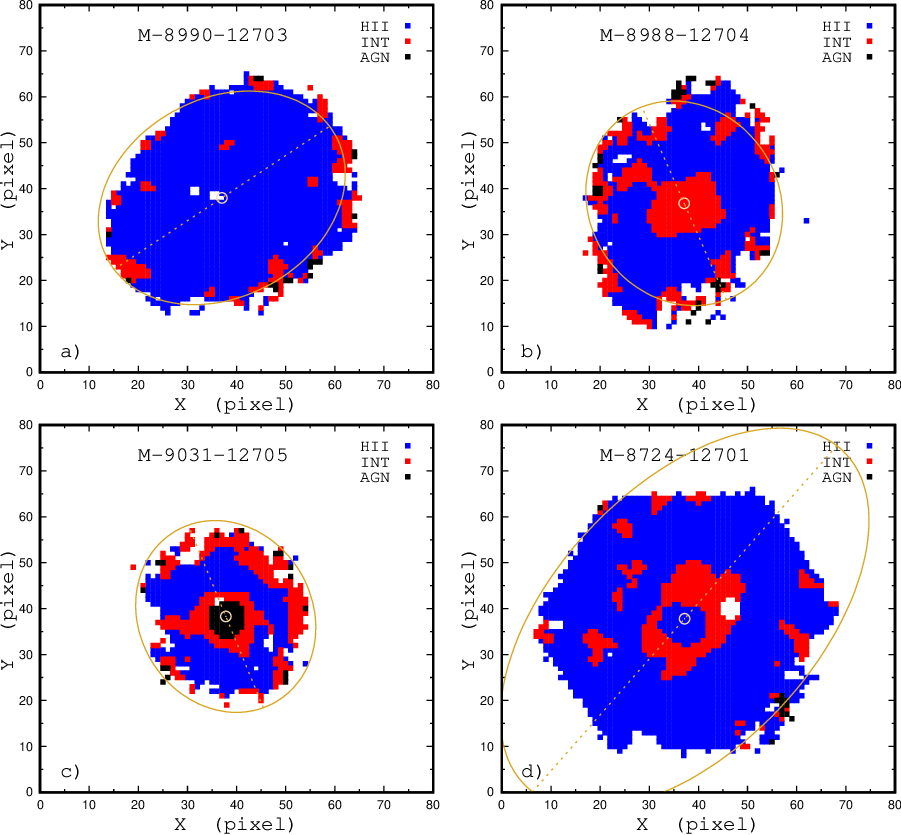}}
\caption{
  Four configurations of the distribution of spaxels with spectra of different BPT types in the circumnuclear regions of cSB galaxies.
  The BPT radiation types for individual spaxels are colour-coded. The yellow circle shows the kinematic centre of the galaxy, the line indicates
  a position of the major kinematic axis of the galaxy, and the ellipse is the optical radius.
}
\label{figure:four-configurations}
\end{figure}

A significant star formation rate at the centre of galaxy implies that  an appreciable  amount of gas is available (flowing on) at the centre of the cSB galaxy.   
It is believed \citep[e.g.][]{Heckman2014} that the nuclei of most (and possibly all) massive galaxies host a central SMBH.
It has been known for a long time that active galactic nuclei (AGNs) are powered by the gas accretion onto SMBHs \citep{Zeldovich1964,Salpeter1964,LyndenBell1969}.
The gas at the centre of galaxy can serve not only as a raw material for the star formation but also as a fuel for the active galactic nucleus. 
Thus, we can expect both main excitation mechanisms (i.e. starburst or AGN) to be found at the centres of cSB galaxies.

The intensities of strong emission lines can be used to separate different types of spaxel spectra according to their main  excitation mechanism (i.e. starburst or AGN). 
A widely used spectral classification of emission-line spectra is the [O\,{\sc iii}]$\lambda$5007/H$\beta$ versus \ [N\,{\sc ii}]$\lambda$6584/H$\alpha$ diagnostic diagram 
suggested by \citet{Baldwin1981}. The  BPT classification diagram can be seen in Fig.~\ref{figure:bpt-example}.  
\citet{Kauffmann2003}  found an empirical demarcation line between the star-forming and AGN spectra in the BPT diagram (solid line in Fig.~\ref{figure:bpt-example}).
This demarcation line  can be interpreted as an upper limit of pure star forming  spectra. The spectra located left (below) the demarcation line of \citet{Kauffmann2003}
are referred to here as the H\,{\sc ii} region-like (or the SF-like) spectra (blue points in Fig.~\ref{figure:bpt-example}).
\citet{Kewley2001} determined a theoretical demarcation line between the star forming  and AGN spectra in the BPT diagram (short-dashed line in Fig.~\ref{figure:bpt-example}).
  This demarcation line is a theoretical upper limit for starburst models on the diagnostic diagram. 
  The spectra located right (above) the demarcation line of \citet{Kewley2001} is referred to 
as the AGN-like spectra  (black points in Fig.~\ref{figure:bpt-example}). The spectra located between the demarcation lines of \citet{Kauffmann2003} and \citet{Kewley2001} are referred to as the intermediate (INT) spectra (red points in Fig.~\ref{figure:bpt-example}). In the literature, those spectra are also referred  to as composite or 
transition spectra \citep[e.g.][]{Davies2014a,Pons2014,Pons2016}. The long-dashed line in Fig.~\ref{figure:bpt-example} is the dividing line between
Seyfert galaxies and low-ionisation nuclear emission line regions (LINERs) defined by \citet{CidFernandes2010}. It should be noted that the exact location of the dividing line
between starbursts (H\,{\sc ii} regions) and AGNs in the BPT diagram is still debatable  \citep[see, e.g.][]{Kewley2001,Kauffmann2003,Stasinska2006,Stasinska2008,Herpich2016}. 

There are four configurations obtained for the distribution of the radiation of different BPT types in the circumnuclear regions of cSB galaxies (see Fig.~\ref{figure:four-configurations}). 
In the first (SF), the central area involves spaxels with the  H\,{\sc ii} region-like radiation only, as seen in panel (a) in Fig.~\ref{figure:four-configurations}. 
In the second, (INT),  the radiation at the centre of a galaxy is the intermediate BPT type,  as seen in panel (b) in Fig.~\ref{figure:four-configurations}. 
In the third (AGN), the innermost region of the AGN-like radiation is surrounded by a ring of radiation of the intermediate type,  as seen in panel (c) in Fig.~\ref{figure:four-configurations}.  
In the fourth (SF+INT), the inner region of the   H\,{\sc ii} region-like radiation is surrounded by a ring of 
radiation of the intermediate type, as seen in panel (d) in Fig.~\ref{figure:four-configurations}.  

%===============    Fig  No 4          SHa-SR3b-HII 
\begin{figure*}
\resizebox{1.00\hsize}{!}{\includegraphics[angle=000]{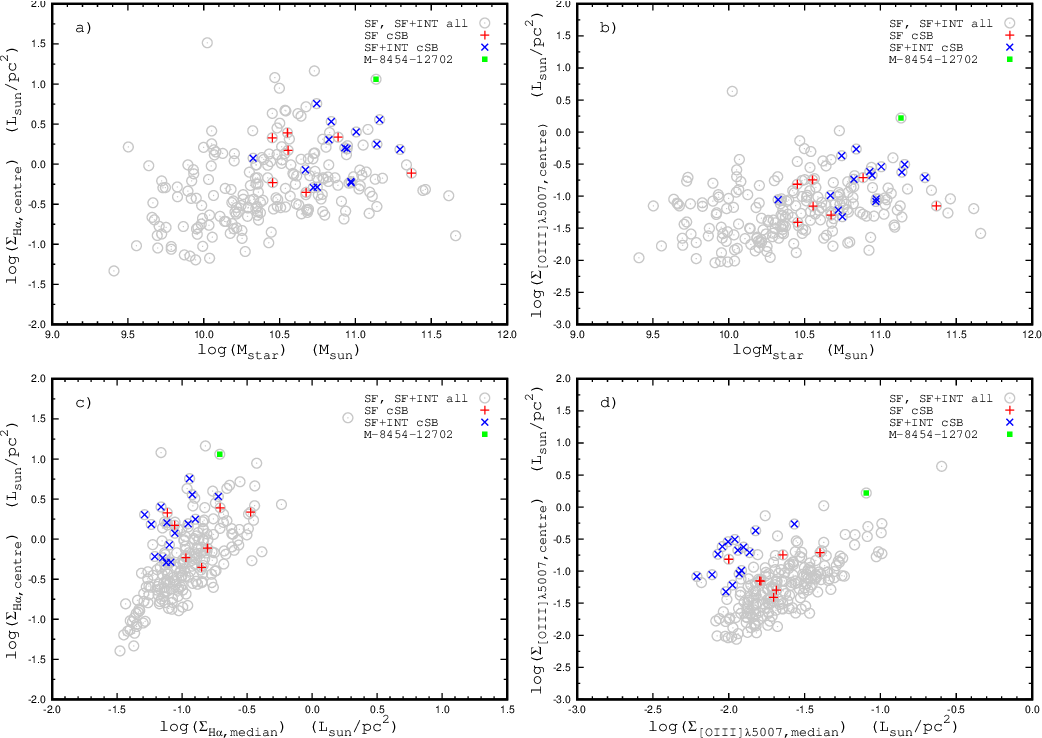}}
\caption{
  Central sufrace brightness of H$\alpha$ and oxygen [O\,{\sc iii}]$\lambda$5007 emission lines in galaxies with the circumnuclear regions of the SF and SF+INT BPT types.
  {\em Panel} {a:} Central surface brightness $\Sigma_{H\alpha, centre}$ as a function of galaxy's stellar mass.
  The grey circles denote all the galaxies, the red plus signs mark the cSB galaxies with the circumnuclear regions of the SF BPT type, and blue crosses show the cSB galaxies with the
    circumnuclear regions of the SF+INT BPT type. The green square is the MaNGA galaxy M-8454-12702.
  {\em Panel} {b:} Same as panel (a) but for the oxygen [O\,{\sc iii}]$\lambda$5007 emission line. 
  {\em Panel} {c:} Central surface brightness $\Sigma_{H\alpha, centre}$ as a function of the median surface brightness $\Sigma_{H\alpha, median}$. The designations are the same
  as in panel (a). 
  {\em Panel} {d:}  Same as panel (c) but for the oxygen [O\,{\sc iii}]$\lambda$5007 emission line. 
}
\label{figure:sha-sr3b-hii}
\end{figure*}

The radiation distributions of different BPT types in the circumnuclear regions of our sample of cSB galaxies are shown in panels of column  (d) in
Figs.~\ref{figure:app-fig1} --  \ref{figure:app-fig5}. Seven cSB galaxies show the SF configuration of the radiation distribution, two cSB galaxies show INT configuration,
the AGN configuration is found in six cSB galaxies, and the SF+INT configuration is revealed in fifteen cSB galaxies. Thus, the SF radiation makes dominant contribution to the
excitation of 22 cSB galaxies (SF and SF+INT configurations), while the AGN radiation dominates in the excitation of 6 cSB galaxies.

  The position of the object on the BPT diagram is a widely used  method of identifying the ionising source of nebular emission. However, the credibility of the  classification
  of the ionising source of
  the region using only the BPT diagram has been questioned \citep[e.g.][]{Sanchez2014,Sanchez2021,Sanchez2024,Lacerda2018,D'Agostino2019}. 
  In particular, it was found that the demarcation line of \citet{Kauffmann2003} in the BPT diagram does not provide the complete selection of the regions ionised by SF.
  From one side, \citet{Sanchez2014} noted that some H\,{\sc ii} regions can be found above the demarcation line of \citet{Kauffmann2003}. 
  From another side, SF can be not the only ionising source in some objects below this demarcation line. \citet{KauffmannHeckman2009} created a set of mixing line
  ‘trajectories’ by averaging the emission-line luminosities of the star-forming and AGN objects in different proportions and found that the object locates below the
  demarcation line of \citet{Kauffmann2003} when the contribution from SF is greater than half of the total  [O\,{\sc iii}] luminosity.

The intermediate spectra of spaxels in the circumnuclear region (e.g. a ring of radiation of the intermediate type around the innermost region of the AGN-like radiation)
can be explained by the AGN-SF mixing where both the SF and AGN radiations make a contribution to the gas excitation \citep{KauffmannHeckman2009,Davies2014a,Davies2014b,Davies2016,Pilyugin2020}.
The proportion of the SF and AGN radiation contributions change smoothly with radius, from the AGN-like sources in spaxel spectra at the galactic centre, to 
the SF-like sources in spaxel spectra at larger galactocentric distances.
Many galaxies involve spaxels of the intermediate spectra located far from the centres of galaxies \citep[e.g.][]{Belfiore2016,Hviding2018,Parkash2019}.
This offers evidence that the intermediate spectra in those spaxels are not produced by AGNs, instead, a source of the ionising photons in those spaxels can be attributed to  
the hot, low-mass evolved (postasymptotic giant branch) stars (HOLMES) \citep{Stasinska2006,Stasinska2008,Sarzi2010,YanBlanton2012,Singh2013}.    
\citet{Belfiore2016} noted that  sources of ionising photons in spaxels of the intermediate spectra near the centres of some galaxies
can be also attributed to the HOLMES.
  \citet{Sanchez2014} noted that some H\,{\sc ii} regions can be found above the demarcation line of \citet{Kauffmann2003}. Thus, the area in the BPT diagram between
  the demarcation curves of  \citet{Kauffmann2003} and \citet{Kewley2001}, as revealed by the intermediate spectra, is populated by the objects related to different ionising sources.

  New methods for distinguishing the ionising source by use of another diagnostic indicators have been considered.  One suggestion was to use the equivalent width of the emission H$\alpha$
  line, $EW_{{\rm H}\alpha}$, \citep{CidFernandes2010,CidFernandes2011,Sanchez2014,Lacerda2018,Sanchez2021,Sanchez2024} and the gas velocity dispersion, $\sigma_{{\rm h}\alpha}$,
  \citep{D'Agostino2019,Johnston2023,Sanchez2024} as diagnostic indicators in addition to the emission-line ratios which are at the base of the BPT diagnostic diagram.  
  Recently, \citet{Sanchez2024} proposed a new method that explores the location of different ionising sources in a diagram, which compares  equivalent width of the emission
  H$\alpha$ line, $EW_{{\rm H}\alpha}$, and gas velocity dispersion, $\sigma_{{\rm h}\alpha}$, (WHaD diagram). These authors defined different areas in which the ionising source could be classified as:
  (1) SF:\ ionisation due to young-massive OB stars, related to recent star-formation activity; (2) sAGNs and wAGNs:\ ionisation due to strong (weak) AGNs, and other sources of ionisation
  such as high-velocity shocks; and (3) Ret: ionisation due to hot old low-mass evolved stars (post-AGBs), associated with retired regions within galaxies (in which there is no star-formation).
  In particular, they classified as SF those sources with a  $EW_{{\rm H}\alpha} >$ 6 {\AA} and  $\sigma_{{\rm H}\alpha} <$ 57 km/s. 

  In the current study, we have estimated the oxygen abundances in all the spaxels with the H\,{\sc ii} region-like spectra according to the BPT classification, that is,
  in the spaxels located below the demarcation line of \citet{Kauffmann2003} in the BPT diagram. We also carried out the WHaD classification. 
  The comparison between the BPT and WHaD diagrams for our sample of the cSB galaxies is shown in Figs.~\ref{figure:appb-fig1}-\ref{figure:appb-fig3} in the appendix.
  The radial distributions of the oxygen abundances  for individual spaxels in the cSB galaxies are shown by grey points in the panels of column  (c) in
  Figs.~\ref{figure:app-fig1}-\ref{figure:app-fig5}. The spaxels that show the H\,{\sc ii} region-like spectra according to the BPT classification, but do not identify
  in the SF area in the WHaD diagram,
  are marked by green plus signs in the panels of column  (c) in Figs.~\ref{figure:app-fig1}-\ref{figure:app-fig5}. The classification according to the WHaD diagram is
  taken into consideration in the discussion of the results.

\subsection{Surface brightness in the emission lines at the centres of cSB galaxies} 
%=====================

%===============    Fig  No 5          M 8454-12702
\begin{figure}
\resizebox{1.00\hsize}{!}{\includegraphics[angle=000]{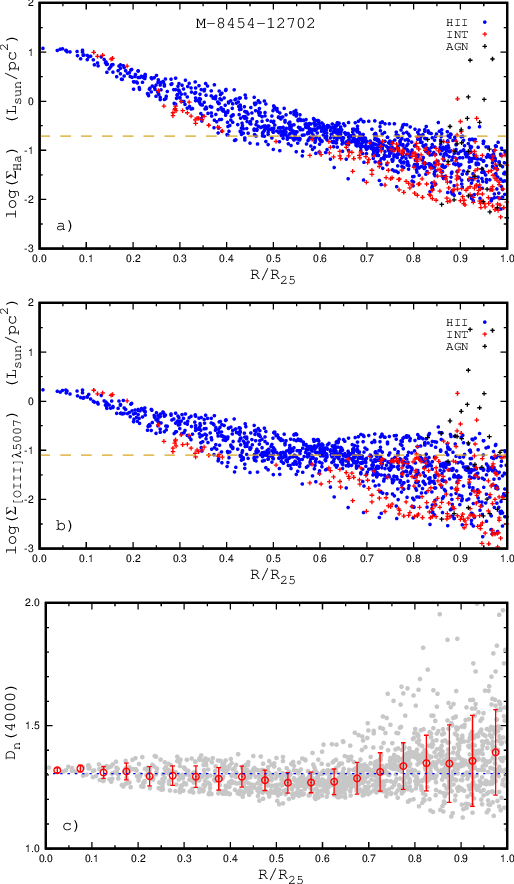}}
\caption{
  Radial disributions of the sufrace brightness in the H$\alpha$ and oxygen [O\,{\sc iii}]$\lambda$5007 emission lines and  D$_{n}$(4000) index in the MaNGA galaxy M-8454-12702.
  {\em Panel} {a:} Surface brightness $\Sigma_{H\alpha}$ as a function of radius. The blue symbols denote the spaxels with the H\,{\sc ii} region-like spectra,
   dark symbols mark the spaxels with the AGN-like spectra, and the red symbols are the spaxels with the intermediate spectra. The golden dashed line shows the
  median value $\Sigma_{H\alpha},median$ of the surface brightness in the spaxels with the optical radius of the galaxy.
  {\em Panel} {b:} Same as panel (a) but for the oxygen [O\,{\sc iii}]$\lambda$5007 emission line. 
  {\em Panel} {c:} D$_{n}$(4000) index  as a function of radius.
  The grey points denote D$_{n}$(4000) indeces in the individual spaxels, red circles mark the median values of the D$_{n}$(4000) in bins of 0.05 in the fractional
  radius $R/R_{25}$,  bars show the scatter in the D$_{n}$(4000) about the median value in the bins, and the blue dashed line shows the median value of the  D$_{n}$(4000) 
  in all the spaxels within optical radius.  
}
\label{figure:m-8454-12702}
\end{figure}

%===============    Fig  No 6          Msp - SFR for total sample and cSB subsample 
\begin{figure}
\resizebox{1.00\hsize}{!}{\includegraphics[angle=000]{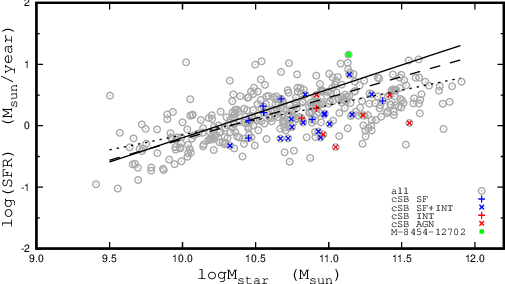}}
\caption{
  Star formation rate as a function of the stellar mass of the galaxy.
  The grey circles denote all the galaxies,  blue plus signs mark the cSB galaxies with the circumnuclear regions of the SF BPT type,  blue crosses show the cSB galaxies with the
  circumnuclear regions of the SF+INT BPT type,  red plus signs are the cSB galaxies with the circumnuclear regions of the INT BPT type, and red crosses show the cSB galaxies with the
  circumnuclear regions of the AGN BPT type.   The green square is the MaNGA galaxy M-8454-12702.
  The SFR -- $M_{\star}$ relations for the MaNGA galaxies obtained by \citet{CanoDiaz2019} are shown by the dashed line for the early-type spirals and by the solid line for the late-type
  spirals.  The dotted line denotes the SFR -- $M_{\star}$ relation for the present-day epoch from \citet{Speagle2014}.
}
\label{figure:m-sfr}
\end{figure}

%===============    Fig  No 7          SHa-SR3b-AGN 
\begin{figure*}
\resizebox{1.00\hsize}{!}{\includegraphics[angle=000]{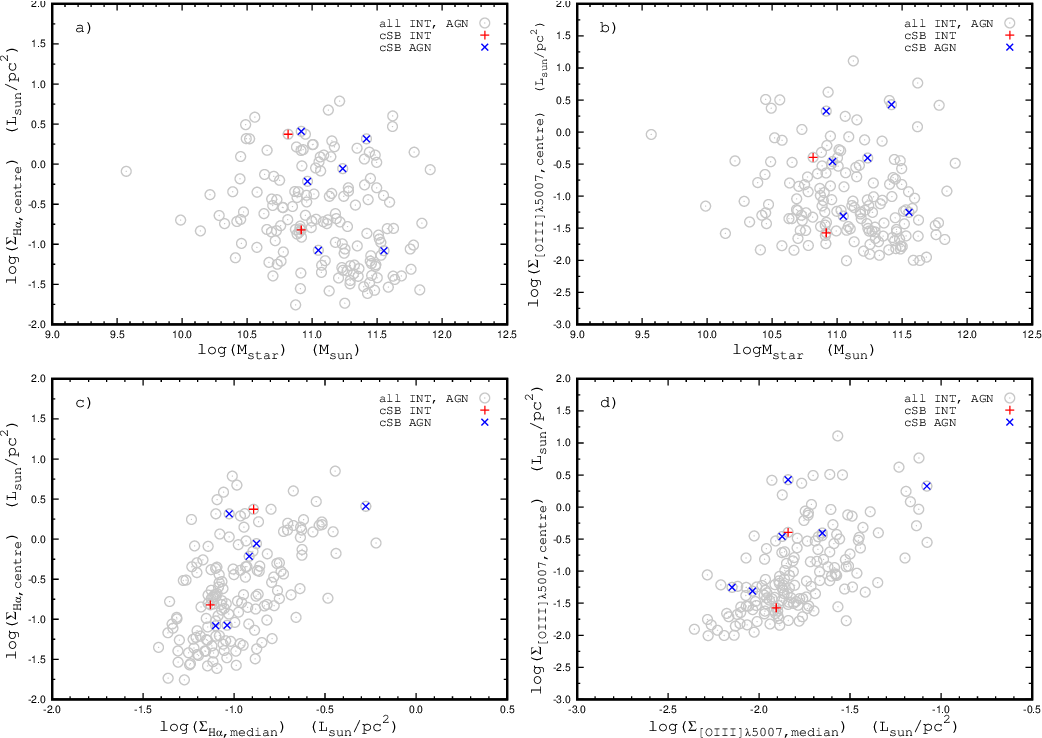}}
\caption{
  Central sufrace brightness in the H$\alpha$ and oxygen [O\,{\sc iii}]$\lambda$5007 emission lines in galaxies with the circumnuclear regions of the INT and AGN BPT types.
  {\em Panel} {a:} Central surface brightness $\Sigma_{H\alpha},centre$ as a function of stellar mass of galaxy.
  The grey circles denote all the galaxies,  red plus signs mark the cSB galaxies with the circumnuclear regions of the INT BPT type,  and blue crosses show the cSB galaxies with the
    circumnuclear regions of the AGN BPT type. 
  {\em Panel} {b:} Same as panel (a) but for the oxygen [O\,{\sc iii}]$\lambda$5007 emission line. 
  {\em Panel} {c:} Central surface brightness $\Sigma_{H\alpha},centre$ as a function of the median surface brightness $\Sigma_{H\alpha},median$. The designations are the same
  as in panel (a). 
  {\em Panel} {d:}  Same as panel (c) but for the oxygen [O\,{\sc iii}]$\lambda$5007 emission line. 
}
\label{figure:sha-sr3b-agn}
\end{figure*}

Here, we compare the surface brightness in the emission lines at the centres of cSB and other galaxies. The observed line fluxes in the spaxel spectra were corrected
for the interstellar reddening and their absolute fluxes are estimated for the adopted distance to the galaxy. The surface brightness in the  H$\alpha$ and [O\,{\sc iii}]$\lambda$5007
emission lines are estimated for each spaxels taking into account the galaxy inclination (reduced to the face-on position). The radial distributions of the surface brightness in the
H$\alpha$ line for our sample of the cSB galaxies are shown in panels of column  (b) in Figs.~\ref{figure:app-fig1}-\ref{figure:app-fig5}. An inspection of those figures shows
that the typical feature of the radial distributions of the H$\alpha$ surface brightness in a cSB galaxy is a well defined central peak.     
The H$\alpha$ flux of a region is a widely used indicator of the current star formation rate \citep{Kennicutt1998}. The central peaks in the H$\alpha$ surface brightness in galaxies
selected on the base of the radial D$_{n}$(4000) index distributions confirm that the starbursts take place at the centres of the cSB galaxies.
It should be noted that the central peaks in the H$\alpha$ surface brightness can be partly attributed to the presence of the AGN-like source at the centre of the galaxy (see discussion below). 
Strictly speaking, the H$\alpha$ flux is an indicator of the current star formation rate in regions with the H\,{\sc ii} region-like radiation. The AGN makes a dominant contribution to
the high H$\alpha$ surface brightness at the centre of the cSB galaxy with the AGN-like radiation in the circumnuclear region.

To compare the surface brightness in the cSB and other galaxies, we specify the radial distribution of the surface brightness in the  H$\alpha$ line in a galaxy
using two values. The median value of the  H$\alpha$ surface brightness in five spaxels nearest to the centre is adopted as the H$\alpha$ surface brightness
at the centre of the galaxy, $\Sigma_{H\alpha,centre}$. We also estimated the median values of the  H$\alpha$ surface brightness for all the spaxels within the optical
radius of the galaxy, $\Sigma_{H\alpha,median}$. In a similar way, we determined the central  $\Sigma_{[OIII]\lambda5007,centre}$ and median $\Sigma_{[OIII]\lambda5007,median}$ surface
brightness in the oxygen emission line  [O\,{\sc iii}]$\lambda$5007. 

Panel (a) of Fig.~\ref{figure:sha-sr3b-hii} shows the central H$\alpha$ surface brightness, $\Sigma_{H\alpha,centre}$, as a function of the stellar mass for the galaxies  with the
H\,{\sc ii} region-like radiation at the centre.  The grey circles denote the galaxies from our total sample,  red plus signs mark the cSB galaxies with the circumnuclear regions
of the SF BPT type, and  blue crosses show the cSB galaxies with the circumnuclear regions of the SF+INT BPT type.
The cSB galaxies of our sample are massive galaxies, log($M_{\star}/M_{\sun}) \ga$ 10.5 (Fig.~\ref{figure:sha-sr3b-hii}). Inspection of panel (a) of Fig.~\ref{figure:sha-sr3b-hii} shows that
the central H$\alpha$ surface brightness, $\Sigma_{H\alpha,centre}$, of cSB galaxies tend to be shifted towards the high surface brightness for galaxies of the given
stellar masses. Consequently, the rates of the star formation of cSB galaxies tend to be shifted towards the high star formation rate for galaxies of a given stellar mass.
However, the star formation rates at the centres of the cSB galaxies are not extraordinary high. Instead, they are comparable to the central star formation rates in other galaxies.
Moreover, the star formation rates at centres of some
other galaxies (e.g. in the galaxy M-8454-12702, marked by the green square in  Fig.~\ref{figure:sha-sr3b-hii}) are significantly higher than that in the cSB galaxies.
In Fig.~\ref{figure:m-8454-12702}, we show the radial distributions of the sufrace brightness in the H$\alpha$ and oxygen [O\,{\sc iii}]$\lambda$5007 emission lines and D$_{n}$(4000) indices in
the MaNGA galaxy M-8454-12702. Panel (c) of Fig.~\ref{figure:m-8454-12702} shows that there is no the local decrease in the D$_{n}$(4000) indices at the centre of   M-8454-12702. Instead,
the D$_{n}$(4000) values are low across the whole galaxy. Therefore, the galaxy M-8454-12702 does not satisfy to our criterion for the selection of the cSB galaxies. 
The surface brightness,  $\Sigma_{H\alpha}$, changes smoothly along the radius without a well defined central peak,  panel (a) of Fig.~\ref{figure:m-8454-12702}.
Thus, the behaviour of $\Sigma_{H\alpha}$ with the radius confirms that the M-8454-12702 is not a galaxy with the central starburst; rather, it is a starburst galaxy.
The surface brightness,  $\Sigma_{[OIII]\lambda5007}$, also changes smoothly along the radius without a well defined central peak,  panel (b) of Fig.~\ref{figure:m-8454-12702}.
It should be noted that the M-8454-12702 is a massive (log($M_{\star}/M_{\sun})$ = 11.229) galaxy.

Figure~\ref{figure:m-sfr} shows the global star formation rate as a function of the galaxy stellar mass.  We estimate the global star formation rate from
the H$\alpha$ luminosity of a galaxy $L_{{\rm H}{\alpha}}$ using the calibration relation of \citet{Kennicutt1998}, reduced by \citet{Brinchmann2004} for the Kroupa initial mass function
\citep{Kroupa2001}:  
\begin{equation}
\log {\rm SFR}  = \log L_{{\rm H}{\alpha}} -41.28 . 
\label{equation:sfr}
\end{equation}
The  H$\alpha$ luminosity of a galaxy, $L_{{\rm H}{\alpha}}$, was determined as a sum of the H$\alpha$ luminosities of the spaxels with  H\,{\sc ii}-region-like spectra within the optical
radius.
  The SFR -- $M_{\star}$ diagram for MaNGA galaxies has been considered by \citet{Sanchez2018,Sanchez2022,CanoDiaz2019}. The  SFR  -- $M_{\star}$ relations separate  for the star-forming
  early-type spiral galaxies (which comprises the galaxies of the S0a, Sa, Sab, and Sb morphological types) and late-type (Sbc, Sc, Scd, Sm, and Irr) spirals were suggested in
  \citet{CanoDiaz2019}. Those relations are shown in Fig.~\ref{figure:m-sfr} by dashed and solid lines.  
\citet{Speagle2014} investigated the evolution of the star-forming galaxy of main sequence in the star formation rate -- stellar mass diagram using a compilation of 25 studies from the
literature. They found the `consensus' relation  SFR = $f$($M_{\star},t$):   
\begin{eqnarray}
       \begin{array}{lll}
\log SFR(M_{\star},t) & = & (0.84 - 0.026 \times t) \log M_{\star} \\
                    & - & (6.51 - 0.11 \times t),
     \end{array}
\label{equation:speagle}
\end{eqnarray}
where $t$ is the age of the Universe in Gyr, stellar mass $M_{\star}$ is in the solar mass unit, and the SFR is in the solar mass per year. Their SFR -- $M_{\star}$ relation for the present-day
epoch ($t$ = 13.6 Gyr) is shown by the dotted line in Fig.~\ref{figure:m-sfr}. {The locations of galaxies of our sample in the SFR -- $M_{\star}$ diagram are in agreement with the relations
  from \citet{CanoDiaz2019} and \citet{Speagle2014}.}  Inspection of Fig.~\ref{figure:m-sfr} shows that the values of the global star formation rate of cSB galaxies tend to be shifted towards
  the lower value of
the star formation rate for galaxies of a given stellar mass in contrast to the behaviour of the cSB galaxies in the central H$\alpha$ surface brightness (central star formation rate)
-- stellar mass diagram in panel (a) of Fig.~\ref{figure:sha-sr3b-hii}.

Panel (b) of Fig.~\ref{figure:sha-sr3b-hii} shows the central surface brightness in the oxygen line, $\Sigma_{[OIII]\lambda5007,centre}$, as a function of the stellar mass for the galaxies  with the
H\,{\sc ii} region-like radiation at the centre. Examination of the $\Sigma_{[OIII]\lambda5007,centre}$ -- $M_{\star}$ diagram (panel (b) of Fig.~\ref{figure:sha-sr3b-hii}) confirms the conclusions
derived from the analysis of the $\Sigma_{H\alpha,centre}$ -- $M_{\star}$ diagram (panel (a) of Fig.~\ref{figure:sha-sr3b-hii}).

Panel (c) of Fig.~\ref{figure:sha-sr3b-hii} shows central H$\alpha$ surface brightness $\Sigma_{H\alpha,centre}$ as a function of the median surface brightness $\Sigma_{H\alpha,median}$ for the
galaxies  with the H\,{\sc ii} region-like radiation at the centre. Inspection of panel (c) of Fig.~\ref{figure:sha-sr3b-hii} shows that the cSB galaxies with
the SF+INT configuration of the radiation in the circumnuclear regions are located in the upper envelope of the band outlined by the other galaxies in the
$\Sigma_{H\alpha,centre}$ -- $\Sigma_{H\alpha,median}$ diagram.  However, such a shift is less evident (if at all present)\ for the cSB galaxies with the SF configuration of the radiation in the circumnuclear regions.
Panel (d) of Fig.~\ref{figure:sha-sr3b-hii} shows the central surface brightness in the oxygen line $\Sigma_{[OIII]\lambda5007,centre}$ as a function of the median surface brightness, 
$\Sigma_{[OIII]\lambda5007,median}$, for the galaxies  with the H\,{\sc ii} region-like radiation at the centre.
An examination of the $\Sigma_{[OIII]\lambda5007,centre}$ --  $\Sigma_{[OIII]\lambda5007,median}$ diagram results in similar conclusions as in the analysis of the
$\Sigma_{H\alpha,centre}$ -- $\Sigma_{H\alpha,median}$ diagram.

Thus, the above considerations suggest the following properties of a typical cSB galaxy  with the H\,{\sc ii} region-like radiation at the centre.  
This is a massive, log($M_{\star}/M_{\sun}) \ga$ 10.5, well evolved galaxy with the reduced, on average, global current star formation rate compared to 
galaxies of a given stellar masses.
The radial distributions of the H$\alpha$ surface brightness (star formation rate) shows a well defined central peak.
The star formation rate at the centre is comparable to the central star formation rates in other galaxies of a given masses but the excess of the star formation rate at the centre
over the median star formation rate across the galaxy is higher than in other galaxies of a given  median star formation rate.

  The spectra of the spaxels at the centres of the cSB galaxies with the circumnuclear regions of the SF+INT BPT type correspond to the H\,{\sc ii} region-like type according
  to the BPT classification. 
  However, those spaxels are of the wAGN (or sAGN in some cases) type, according to the WHaD classification (Figs.~\ref{figure:appb-fig1}-\ref{figure:appb-fig3}) because of 
  the values of the gas velocity dispersion, $\sigma_{{\rm H}\alpha}$, in  spaxels exceeding 57 km/s, which is the threshold value for the SF object in the WHaD classification \citep{Sanchez2024}.
  This suggests that both the SF and AGN-like radiation can make a contributions to the excitation of gas in those spaxels. We consider whether the shift of the central surface brightness
  in the oxygen line, $\Sigma_{[OIII]\lambda5007,centre}$, towards the upper envelope of the band in the $\Sigma_{[OIII]\lambda5007,centre}$ --  $\Sigma_{[OIII]\lambda5007,median}$
  (and $\Sigma_{H\alpha,centre}$ -- $\Sigma_{H\alpha,median}$) diagram can be attributed to the contribution of the AGN-like source.  Taking into account that the spectrum is of the SF type according 
  to the BPT classification (i.e. object is located below the demarcation line of \citet{Kauffmann2003}) when the contribution from SF is greater than half of the total  [O\,{\sc iii}]
  luminosity \citep{KauffmannHeckman2009}; thus, we can expect that the shift caused by the contribution of the AGN-like source is less than 0.3 dex. A close examination of panel (d) of
  Fig.~\ref{figure:sha-sr3b-hii} shows that the total shifts of the  central surface brightness in the oxygen line $\Sigma_{[OIII]\lambda5007,centre}$ of the cSB galaxies with the
  circumnuclear regions of the SF+INT BPT type towards the upper envelope of the band in the $\Sigma_{[OIII]\lambda5007,centre}$ --  $\Sigma_{[OIII]\lambda5007,median}$ diagram can be significantly
  larger than 0.3 dex and, consequently, only a part of the shift (but not the total shift) can be attributed to the contribution of the AGN-like source.

Figure~\ref{figure:sha-sr3b-agn} shows the $\Sigma_{H\alpha,centre}$ -- $M_{\star}$ diagram, panel (a), the $\Sigma_{[OIII]\lambda5007,centre}$ -- $M_{\star}$ diagram, panel (b),
the $\Sigma_{H\alpha,centre}$ -- $\Sigma_{H\alpha,median}$ diagram, panel (c), and the $\Sigma_{[OIII]\lambda5007,centre}$ --  $\Sigma_{[OIII]\lambda5007,median}$ diagram, panel (d),
for galaxies with the AGN-like and intermediate radiation in the circumnuclear regions.
An inspection of Fig.~\ref{figure:sha-sr3b-agn} shows that the shift of the central surface brightness in the cSB galaxies towards the higher values is noticeable in the
$\Sigma_{[OIII]\lambda5007,centre}$ --  $\Sigma_{[OIII]\lambda5007,median}$ diagram only, as seen in panel (d) of Fig.~\ref{figure:sha-sr3b-agn}.

%===============    Fig  No 8          Repeated 
\begin{figure*}
\resizebox{1.00\hsize}{!}{\includegraphics[angle=000]{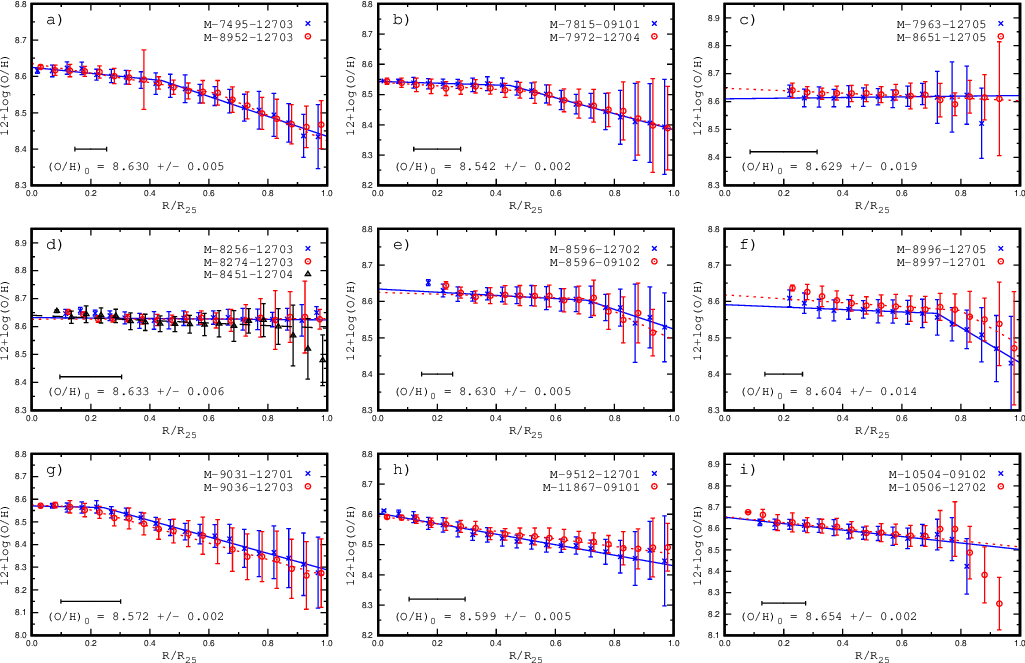}}
\caption{
  Comparison of the radial distributions of oxygen abundances in the galaxy determined from the independent MaNGA observations of the same galaxy.
  In each panel, the symbols denote the median values of the O/H in bins of 0.05 in the fractional radius, $R/R_{25}$, and the bars show the scatter in O/H
  about the median value in the bins. The lines show the relations we determined for the radial abundance distributions. The horizontal bar indicates
    the point spread function in the MaNGA measurements.
}
\label{figure:repeated}
\end{figure*}

The properties of the cSB galaxies with the AGN and INT types of the radiation in circumnuclear regions are somewhat controversial. From one side, the decrease in the D$_{n}$(4000) indices
(stellar ages) at the centre of such galaxies offers evidence on the central starburst. From another side, the AGN (but not the starburst) is a dominant source of the ionising radiation
in the circumnuclear region. This can reveal evidence that the central starburst in such galaxy occurred some time ago and is faded this time; therefore the star radiation
makes a small, (and even no) contribution to the ionising radiation in the circumnuclear region.

\subsection{Central oxygen abundances in MaNGA galaxies} 
%=====================

%===============    Fig  No 9    OHo - OHsdss
\begin{figure}
\resizebox{1.00\hsize}{!}{\includegraphics[angle=000]{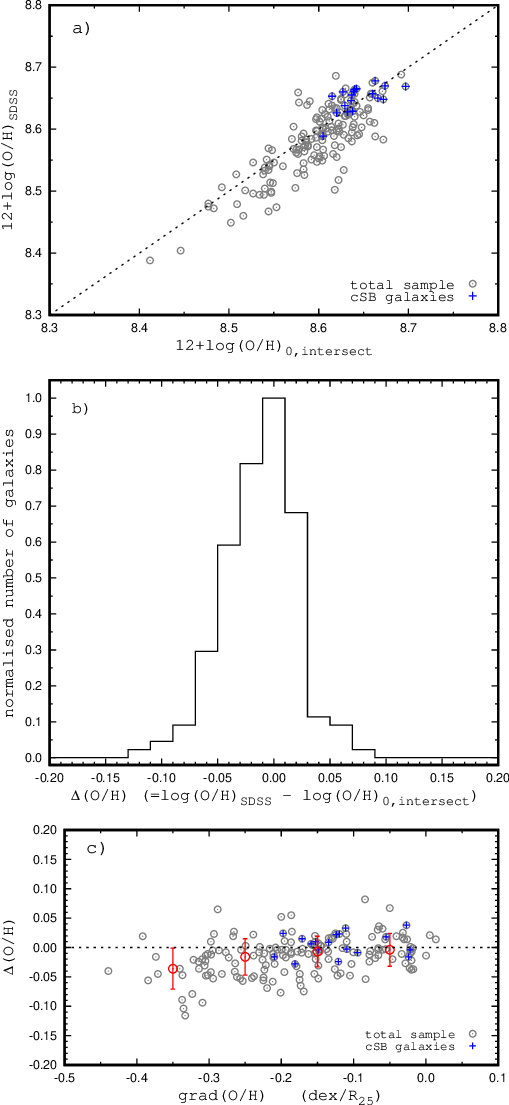}}
\caption{
  Comparison the central intersect oxygen abundances estimated from the metallicity gradient determined on the base of the IFU spectroscopy from the MaNGA, (O/H)$_{0,intersect}$,
  and the local central oxygen abundances obtained from the single-fibre observations from the SDSS, (O/H)$_{SDSS}$.
  {\sl Panel} {a:}  (O/H)$_{SDSS}$ as a function of the  (O/H)$_{0,intersect}$. The grey circles denote all the galaxies, the red plus signs mark the cSB galaxies, and the line is
  one-to-one correspodance.
  {\sl Panel {b}:} Normalised histogram of the differences $\Delta$(O/H) = log(O/H)$_{SDSS}$ -- log(O/H)$_{0,intersect}$ for our sample of galaxies. 
  {\sl Panel {c}:} Difference $\Delta$(O/H) as a function of the gradient slope.
  The grey circles denote all the galaxies, the red plus signs mark the cSB galaxies, 
  The red circles denote the median values of the $\Delta$O/H in bins of 0.1 in the gradient slope, and the bars show the scatter in the $\Delta$O/H 
  about the median value in the bins. 
}
\label{figure:oho-ohsdss}
\end{figure}

%===============    Fig  No  10         OHo - dMpc 
\begin{figure}
\resizebox{1.00\hsize}{!}{\includegraphics[angle=000]{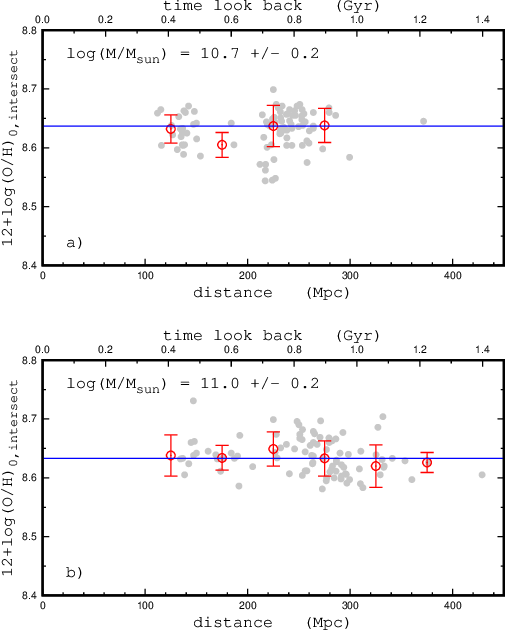}}
\caption{
  Central intersect oxygen abundance (O/H)$_{0,intersect}$ as a function of the distance (look back time).
  {\sl Panel} {a:} Central intersect oxygen abundance as a function of the distance (look back time) for galaxies with masses  10.5 $<$ log($M_{\star}/M_{\sun}$) $<$ 10.9.
  The grey circles denote the abundances in individual galaxies, the red circles mark the median values of the abundances for galaxies in bins of 50 Mpc in the distance,
  and the bars show the scatter in the abundances about the median value in the bins. The line is the median value of abundances for all galaxies from this mass interval.
  {\sl Panel {b}:} The same as panel (a) but for galaxies with masses 10.9 $<$ log($M_{\star}/M_{\sun}$) $<$ 11.3.
}
\label{figure:dmpc-oho}
\end{figure}

We have estimated the oxygen abundances in the spaxels with the H\,{\sc ii} region-like spectra using the $R$ calibration from \citet{Pilyugin2016}. The radial distributions of
the oxygen abundances  for individual spaxels in the cSB galaxies are shown by grey points in the panels of column  (c) in Figs.~\ref{figure:app-fig1}-\ref{figure:app-fig5}. 
In order to minimise the influence of the spaxels with unreliable abundances in determining the radial abundance distribution, we did not use the abundances in individual spaxels,
but the median values of the abundances in bins of 0.05 in $R/R_{25}$ (the red circles in the panels of column  (c) in Figs.~\ref{figure:app-fig1}-\ref{figure:app-fig5}.) 
The majority of the binned oxygen abundances follow well the general metallicity gradient in the disc. Yet, an appreciable deviation from the general trend can take place in bins 
near the centre (e.g. panel (c12) in Fig.~\ref{figure:app-fig2}, panel (c16) in Fig.~\ref{figure:app-fig3}, and panel (c27) in Fig.~\ref{figure:app-fig5}), near the isophotal radius
of the galaxy (e.g. panel (c3) in Fig.~\ref{figure:app-fig1}, panel (c12) in Fig.~\ref{figure:app-fig2}, and panel (c19) in Fig.~\ref{figure:app-fig3}), and near the transition from
the zone of the  H\,{\sc ii} region-like radiation to the zone of the AGN(INT)-like  radiation (e.g. panels (c8) and (c9) in Fig.~\ref{figure:app-fig2}, panel (c30)
in Fig.~\ref{figure:app-fig5}). The deviations of those bins from the general metallicity gradient can be attributed to the uncertainties
in the oxygen abundance determinations, and those bins are rejected in the determinations of the radial abundance gradients. 
The gradients in the galaxies were approximated by a single or broken linear relation (shown by the lines in  panels of column  (c) in Figs.~\ref{figure:app-fig1}-\ref{figure:app-fig5}
for cSB galaxies) and the central oxygen abundance in a galaxy is estimated as an intersect value of the determined radial abundance distribution.

Two independent MaNGA observations are available for eight galaxies from our total sample and three independent MaNGA  observations (M-8274-12703, M-8256-12703, and M-8451-12704)
are available for one galaxy. A comparison between values of the central intersect oxygen abundances derived from different observations of the same galaxy  
provides a possibility to estimate uncertainties in the central intersect oxygen abundances obtained from the MaNGA measurements. 
Fig.~\ref{figure:repeated} shows the comparison of the radial oxygen abundance distributions determined from independent MaNGA observations of the same galaxy for nine galaxies.
The mean value of the central oxygen abundances determined from independent measurements can be considered as `true' central oxygen abundance and the difference between the
value of the central abundance determined from the individual observation and the mean value can be considered as the uncertainty in the central oxygen abundance determination from
the MaNGA measurements. The uncertainties in the central oxygen abundance determination are within 0.01~dex for seven galaxies and are between 0.01 and 0.02~dex for two galaxies,
Fig.~\ref{figure:repeated}.

  The obtained uncertainties in the central oxygen abundance specify the uncertainty (robustness) of the abundances determined   through the used calibration rather than the absolute
  uncertainties. It is known that there are large systematic discrepancies between the abundance values produced by different published calibrations \citep[e.g. see Figure 29 in][]{Sanchez2022}.
  At the present time there is no commonly accepted absolute scale for metallicities of  H\,{\sc ii} regions. It should be noted that the oxygen abundances 
  determined through the $R$ calibration are compatible to the metallicity scale of  H\,{\sc ii} regions defined by the  H\,{\sc ii} regions with abundances obtained through the
  direct $T_{e}$ method. All the abundances considered in the current study are obtained using the same calibration what justifies to make conclusion about agreement (disagreement)  between them.   

It was noted above that if the oxygen abundances in the bins near the centre show the deviations from the general radial abundance gradient then those bins are rejected in the
determination of the radial abundance gradient. \citet{Belfiore2017} noted that the observed metallicities in the spaxels near the centre of a galaxy can be affected by the
‘beam-smearing’ effect of the  point spread function (PSF). The point spread function of the MaNGA measurements is estimated to have a full width at half maximum of 2.5~arcsec
or five pixels \citep{Bundy2015,Belfiore2017}. For low $R_{25}$/PSF ratios and high galaxy inclinations, the error in the oxygen abundance at the centre of a galaxy can be as
large as $\sim$0.04~dex. On the other hand, it cannot be excluded that the deviation of the central oxygen abundance from the general radial gradient can be real under some
conditions. An appreciable episode of star formation occurred at the present-day or at the recent past in the centres of the cSB galaxies. If a starburst alters the (local) oxygen abundance
at the centre of spiral galaxy, then the intersect oxygen abundance does not correspond to the real central abundance in the cSB galaxy.

  The consideration of the alternative classification of the spaxel spectra using the WHaD diagram of \citet{Sanchez2024} can shed light on the origin of the deviation of the spaxels
  from the general gradient. The radial distributions of the oxygen abundances  for individual spaxels in the cSB galaxies are shown by grey points in the panels of column  (c) in
  Figs.~\ref{figure:app-fig1}-\ref{figure:app-fig5}. The spaxels which show the H\,{\sc ii} region-like spectra according to the BPT classification but not locate in the SF area
  in the WHaD diagram (for the brevity sake we refer to those spectra as non-confirmed  H\,{\sc ii} region-like spectra)  are marked by green plus signs. Examination of panels of
  column  (c) in Figs.~\ref{figure:app-fig1}-\ref{figure:app-fig5} shows that the binned oxygen abundance deviates from the general gradient if the bin involves
  spaxels with non-confirmed  H\,{\sc ii} region-like spectra. This justifies the rejection of the binned oxygen abundances with large deviations in the determination of the radial
  abundance gradient in the galaxy.

  However, not all the spaxels with the non-confirmed H\,{\sc ii} region-like spectra demonstrate deviations from the general   metallicity gradient (panels of   column  (c) in
  Figs.~\ref{figure:app-fig1}-\ref{figure:app-fig5}). The following  fact should also be noted, namely, if the deviation of abundance in the spaxel with the non-confirmed
  H\,{\sc ii} region-like spectrum from the general metallicity gradient is caused by the contribution of the AGN-like radiation to the ionisation then we can expect the maximum
  deviations of the abundances in the
  spaxels at the centre of the galaxy. It is not the case for some galaxies. For example, in the galaxy M-8569-12701, the maximum deviations of the oxygen abundances from the general
  trend show the spaxels at radii 0.2 $\la R/R_{25} \la$ 0.3, while the oxygen abundances in the spaxels at $R/R_{25} \la$ 0.2 are close to the general trend (panel (c08) in
  Fig.~\ref{figure:app-fig2}). Similarly, in the galaxy M-8724-12701, the maximum deviations of the oxygen abundances from the general
  trend show the spaxels at radii 0.1 $\la R/R_{25} \la$ 0.2, while the oxygen abundances in the spaxels at $R/R_{25} \la$ 0.1 are close to the general trend (panel (c09) in
  Fig.~\ref{figure:app-fig2}). Thus, the use of the WHaD classification in addition to the BPT classification allows us to reject the spaxels with suspicious abundances (with a deviations
  from the general radial metallicity  gradient). However, we cannot exclude that a fraction of the spaxels with a realistic abundances is also rejected in this way.
  It should be emphasised that the spectra at the centres of the cSB galaxies with the circumnuclear regions of the SF BPT type are  H\,{\sc ii} region-like accoording of both the
  BPT and the WHaD classifications, and the spectra at the centres of the cSB galaxies with the circumnuclear regions of the AGN BPT type are AGN-like type accoording of both
  the BPT and WHaD classifications (Figs.~\ref{figure:appb-fig1}-\ref{figure:appb-fig3}).  The non-confirmed H\,{\sc ii} region-like spectra appear at the centres of
  the cSB galaxies with the circumnuclear regions of the SF+INT BPT type only.

A single-fibre spectroscopy of centres in a large amount of galaxies (including the MaNGA galaxies)  was carried out within the  Sloan Digital Sky Survey (SDSS, \citet{York2000}).
  This provides an independent estimation of the oxygen abundances. The comparison between two independent estimations of the oxygen abundance (the intersect values of the central
  abundances based on the MaNGA measurements and oxygen abundances estimated from the SDSS measurements) can tell us something about the validity of the obtained central abundances
  in MaNGA galaxies and about the agreement of the oxygen abundance at the centre of spiral galaxy with the general radial metallicity gradient.
 The  H\,{\sc ii} region-like SDSS spectra with measurements of all the emission lines necessary for the oxygen abundance determination are available
for the centres of 166 galaxies of our sample. The emission line measurements were taken from the table titled `galSpecLine'\footnote{https://skyserver.sdss.org/dr18/VisualTools/explore/}. 
The SDSS-based oxygen abundances were estimated using the $R$ calibration from \citet{Pilyugin2016}.

Panel (a) of Fig.~\ref{figure:oho-ohsdss} shows the comparison between the central intersect oxygen abundances estimated from the metallicity gradient determined on the base of the IFU
spectroscopy from the MaNGA, (O/H)$_{0,intersect}$,  and the local central oxygen abundances obtained from the single-fibre observations from the SDSS, (O/H)$_{SDSS}$.
Panel (b) of Fig.~\ref{figure:oho-ohsdss} shows the normalised histogram of differences $\Delta$(O/H) = log(O/H)$_{SDSS}$ -- log(O/H)$_{0,intersect}$ for our sample of galaxies. 
The uncertainty of the oxygen abundance estimated through the  $R$ calibration from the individual measurement is within $\sim$0.1~dex \citep{Pilyugin2016}.
The uncertainties in the central intersect oxygen abundance determinations are within $\sim$0.02~dex, Fig.~\ref{figure:repeated}. Inspection of panel (a) and (b) of
Fig.~\ref{figure:oho-ohsdss} shows the differences $\Delta$(O/H) are within those uncertainties.
However, the differences are not perfectly random, there is some excess of the negative values of the differences (the (O/H)$_{SDSS}$ is lower than (O/H)$_{0,intersect}$).

We can assume that this asymmetry in the  $\Delta$(O/H) distribution can be caused by the aperture effect. The SDSS spectra are obtained through 3-arcsec-diameter fibres.
The angular diameters of the bulk of the MaNGA galaxies of our sample lie  within the diapason from 20 to 30 arcseconds. The SDSS fibre covers up to 10-15\% of the galaxy radius (for the
face-on case). The SDSS fibre spectrum involves not only the radiation from the very centre but also the radiation of the circumnuclear region of radius of 0.1--0.15$R_{25}$.
For a galaxy with a steep abundance gradient, the oxygen abundance estimated from the SDSS fibre spectrum will be lower than the oxygen abundance at the centre of the galaxy.
If this is the case then the negative values of the $\Delta$(O/H) should be found in galaxies with the steep abundance gradients. Panel (c) of Fig.~\ref{figure:oho-ohsdss} shows
the $\Delta$(O/H) as a function of the slope of the radial oxygen abundance gradient in the galaxy.  Close examination of panel (c) of Fig.~\ref{figure:oho-ohsdss} shows that
the  $\Delta$(O/H) value correlates with the slope of the gradient in the galaxy, the median value of the  $\Delta$(O/H) changes from $-0.004$~dex for galaxies of the gradients
in the diapason from $-0.1$ to 0.0~dex/$_{R25}$ to $-0.036$~dex for galaxies of the gradients in the diapason from $-0.4$ to $-0.3$~dex/$_{R25}$. Thus, the asymmetry in the  $\Delta$(O/H)
distribution can be attributed to the aperture effect. 

The locations of the cSB galaxies in the  $\Delta$(O/H) -- $M_{\star}$ diagram (panel (a) of Fig.~\ref{figure:oho-ohsdss}) and in the  $\Delta$(O/H) -- grad(O/H) diagram
(panel (c) of Fig.~\ref{figure:oho-ohsdss}) are marked by the red plus signs. The median value of  $\Delta$(O/H) for the cSB galaxies is 0.006~dex. Thus, the median value of the
deviations of the local central abundance from the general metallicity gradient in the cSB galaxies is negligibly small, within the uncertainties in the intersect oxygen abundance
determinations.

The negligibly small local enrichment of the centres of the cSB galaxies in oxygen by the starbursts can be indirectly verified by the following consideration.
The distances to galaxies in our sample lie in the diapason from $\sim$120~Mpc to  $\sim$350~Mpc, depending on the value of stellar mass. This provides a possibility to
estimate the enrichment in oxygen at the centres of galaxies for 0.5 -- 0.8~Gyr. Figure~\ref{figure:dmpc-oho} shows the central oxygen abundance (O/H)$_{0,intersect}$ as a function of
the distance (look back time) for galaxies of the masses from two intervals of stellar masses.
The grey circles in each panel of  Fig.~\ref{figure:dmpc-oho} denote the abundances in individual galaxies, the red circles mark the median values of the abundances for galaxies in
the distance bins of 50~Mpc. The line is the median value of abundances for all galaxies from this mass interval. An inspection of  Fig.~\ref{figure:dmpc-oho} shows that there is
no a systematic variation of the oxygen abundance in massive galaxies for 0.5 -- 0.8~Gyr. This means that the contribution of the star formation at the current epoch in the centres of
massive galaxies to the oxygen enrichment is negligibly low. This is in agreement with the conclusion obtained from the analysis of the N/O - O/H diagram for a sample of well measured
MaNGA galaxies \citep{Pilyugin2024}.

  The chemical enrichment history of galaxies of different masses were investigated using two strategies. First, the oxygen abundance evolution with redshift of the galaxies has
  been considered \citep[e.g.][]{Thuan2010,Moustakas2011,Pilyugin2011,Rodrigues2016}.
  Second, the chemical enrichment history in the galaxy (or in a region of the galaxy) can be determined through the fitting of the observed spectrum by the spectra of a set of
  simple stellar populations of different ages and metallicities. Using this approach, the chemical enrichment histories were determined in the CALIFA  (the Calar Alto Legacy
  Integral Field Area (CALIFA) survey \citep{Sanchez2012,Sanchez2016,GarciaBenito2015}) galaxies  \citep{CampsFarina2021}, and in the MaNGA galaxies  \citep{CampsFarina2022}.
  It has been found in all those studies that there is no an appreciable oxygen enrichment in the centres of massive galaxies (log$M_{\star}/M_{\sun} \ga$ 10.5) for last $\sim$3 Gyrs.
  Our conclusion is in line with this result.

Thus, the central intersect oxygen abundances estimated from the metallicity gradient determined on a base of the IFU spectroscopy from the MaNGA and the local central oxygen abundances
obtained from the single-fibre observations from the SDSS agrees within uncertainties for our sample of galaxies. In particular, the deviations of the local central abundance from the
general metallicity gradient in the cSB galaxies are negligible small, if any. A contribution of the current starbursts in the centres of massive galaxies to the oxygen enrichment is low. 

\section{Conclusions} 
%=====================

We examined whether there are deviations of the local central oxygen abundance in spiral galaxies from the general metallicity gradient giving a special attention to the
galaxies with central starbursts. We have constructed the maps and radial distributions of different characteristics in a sample of 381 MaNGA galaxies using the publicly
available spectral measurements. We also selected a sample of 30 galaxies with central starbursts (cSB galaxies) among galaxies of our sample. The decrease in the D$_{n}$(4000)
index (indicator of the stellar age) in the circumnuclear region is used as the criterion to select the cSB galaxies.

The oxygen abundances in the spaxels with the H\,{\sc ii} region-like spectra {according to the standard BPT classification} were estimated using
the $R$ calibration from \citet{Pilyugin2016}. The radial distribution of
binned oxygen abundances (the median values of the abundances in bins of 0.05~dex in $R/R_{25}$) was approximated by the single or broken linear relation and the central intersect
oxygen abundance was determined. The bins that show a deviation from the general metallicity gradients (in particular, the bins near the centre of the galaxy) were rejected in the
gradient determinations. {Those bins involve the spaxels whose spectra are H\,{\sc ii} region-like according to the standard BPT classification
but do not locate in the area of the SF regions in the WHaD diagram, an alternative classification suggested recently by \citet{Sanchez2024}} 
We estimated the local central oxygen abundances in 166 galaxies of our sample using the single-fibre spectral  observations from the SDSS. We found that 
the central intersect abundances estimated from the metallicity gradient determined on the base of the IFU spectroscopy from the MaNGA and the local central abundances obtained
from the single-fibre observations from the SDSS agree within uncertainties for our sample of galaxies. In particular, the deviations of the local central abundance from the
general metallicity gradient in the cSB galaxies are negligibly small, if any.

The central sturburst implies that  an appreciable  amount of gas presents at the centre of galaxy. The gas at the centre of galaxy  can serve not only as the raw material for
the star formation, but also as the fuel to the active galactic nucleus (AGN). We found that the AGN is the main source of the ionising radiation in the circumnuclear regions of
six cSB galaxies.

The cSB galaxies of our sample with the H\,{\sc ii} region-like radiation at the centre show the following properties: the cSB galaxy is a massive, log($M_{\star}/M_{\sun}) \ga$
10.5, well evolved galaxy with reduced (on average) global current star formation rate in comparison to galaxies of similar stellar masses. The radial distribution of the H$\alpha$ surface
brightness shows a well defined central peak. {Part of the central surface brightness enhancement can be attributed to the contribution of the AGN-like source but the
star formation is the main sourse of the ionising radiation.} The star formation rate at the centre of cSB galaxy is comparable to the central star formation rates in other
galaxies of a given masses, but the excess of the star formation rate at the centre over the median star formation rate across the galaxy is higher than in other galaxies due to
the reduced global star formation rates in the cSB galaxies. The contribution of the current starburst in the centre of cSB galaxy to the oxygen enrichment is negligibly small. 

The properties of the cSB galaxies with the AGN (and INT) types of the radiation in circumnuclear regions are somewthat controversial. From one side, the decrease in the D$_{n}$(4000) 
indices at the centre of such galaxy offers evidence on the central starburst. From the other side, the AGN is the main source of the ionising radiation in the circumnuclear region.
This may serve as evidence that the central starburst in such galaxy is on the faded stage (or has ended) and therefore the star radiation makes a small (if any at all) contribution to the
ionising radiation in the circumnuclear region.

There is no any distinctive link of the central starburst in the galaxy to its present-day environment or to the presence of the bar. This suggests that either there is
another reason responsible for the central starburst or that the central starbursts are not uniform in their origin and may instead originate from different pathways.

%==========================
\begin{acknowledgements}
We are grateful to the anonymous referee for his/her constructive comments. \\ 
L.S.P acknowledges support from the Research Council of Lithuania (LMTLT) (grant no. P-LU-PAR-23-28). \\
This research has made use of the NASA/IPAC Extragalactic Database (NED), which
is funded by the National Aeronautics and Space Administration and operated by
the California Institute of Technology.  \\
We acknowledge the usage of the HyperLeda database (http://leda.univ-lyon1.fr). \\
Funding for SDSS-III has been provided by the Alfred P. Sloan Foundation,
the Participating Institutions, the National Science Foundation,
and the U.S. Department of Energy Office of Science.
The SDSS-III web site is http://www.sdss3.org/. \\
Funding for the Sloan Digital Sky Survey IV has been provided by the
Alfred P. Sloan Foundation, the U.S. Department of Energy Office of Science,
and the Participating Institutions. SDSS-IV acknowledges
support and resources from the Center for High-Performance Computing at
the University of Utah. The SDSS web site is www.sdss.org. \\
SDSS-IV is managed by the Astrophysical Research Consortium for the 
Participating Institutions of the SDSS Collaboration including the 
Brazilian Participation Group, the Carnegie Institution for Science, 
Carnegie Mellon University, the Chilean Participation Group,
the French Participation Group, Harvard-Smithsonian Center for Astrophysics, 
Instituto de Astrof\'isica de Canarias, The Johns Hopkins University, 
Kavli Institute for the Physics and Mathematics of the Universe (IPMU) / 
University of Tokyo, Lawrence Berkeley National Laboratory, 
Leibniz Institut f\"ur Astrophysik Potsdam (AIP),  
Max-Planck-Institut f\"ur Astronomie (MPIA Heidelberg), 
Max-Planck-Institut f\"ur Astrophysik (MPA Garching), 
Max-Planck-Institut f\"ur Extraterrestrische Physik (MPE), 
National Astronomical Observatories of China, New Mexico State University, 
New York University, University of Notre Dame, 
Observat\'ario Nacional / MCTI, The Ohio State University, 
Pennsylvania State University, Shanghai Astronomical Observatory, 
United Kingdom Participation Group,
Universidad Nacional Aut\'onoma de M\'exico, University of Arizona, 
University of Colorado Boulder, University of Oxford, University of Portsmouth, 
University of Utah, University of Virginia, University of Washington, University of Wisconsin, 
Vanderbilt University, and Yale University.
\end{acknowledgements}

%\begin{appendix}
\appendix
%=========

\section{Characteristics of our sample of cSB galaxies}
%=====================

The figures show the distributions of characteristics in the investigated cSB galaxies.
Each galaxy is presented in four panels. 
Panel (a) shows the behaviour of the index D$_{n}$(4000) (which is an indicator of stellar age) along the radius of the galaxy. 
The grey points denote values of the D$_{n}$(4000) index in the individual spaxels, the red circles mark the median values of the D$_{n}$(4000) index 
in bins of 0.05 in the fractional radius $R/R_{25}$, the bars show the scatter in the  values of the D$_{n}$(4000) index about the median value in the bins, and the blue dashed line shows
the median value of the D$_{n}$(4000) index in all the spaxels within the optical radius.  
Panel (b) presents the surface brightness in the emission H$\alpha$ line as a function of radius. The de-reddened and corrected for the galaxy inclination surface brightness
are in units of $L_{\sun}$/pc$^{2}$. The spaxels with the  H\,{\sc ii} region-like spectra are shown by the blue circles, the spaxels with AGN-like spectra are denoted by the dark
crosses, and the spaxels with the intermediate (INT) spectra are marked by the red plus signs. The dashed line shows the median value of the  surface brightness in the emission
H$\alpha$ line in all the spaxels within optical radius.
Panel (c) shows the radial distribution of the oxygen abundance. The grey points denote oxygen abundances in the individual spaxels. The red circles mark the median values
of the O/H in bins of 0.05 in the fractional radius $R/R_{25}$, and the bars show the scatter in O/H  about the median value in the bins. The lines show the relations we
determined for the radial abundance distributions.
{The spaxels which show the H\,{\sc ii} region-like spectra according to the BPT classification but not locate in the SF area in the WHaD diagram
  are marked by green plus signs.
}  
Panel (d) shows the distribution of spaxels of different BPT type spectra across the image of galaxy. The BPT radiation types for individual spaxels are colour-coded.
The yellow circle shows the kinematic centre of the galaxy, the line indicates a position of the major kinematic axis of the galaxy, and the ellipse is the optical radius. 

%===============    Fig  No App 1        
\begin{figure*}
\resizebox{1.00\hsize}{!}{\includegraphics[angle=000]{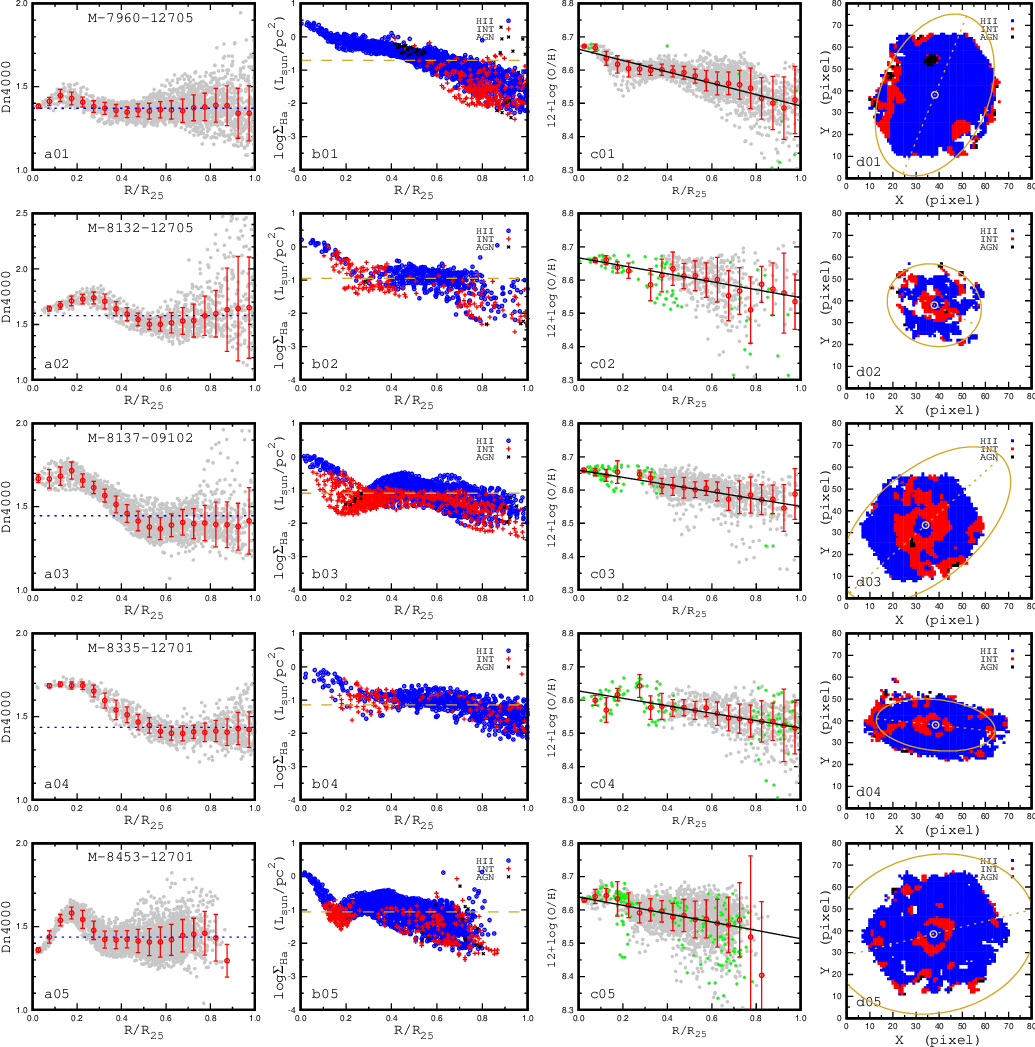}}
\caption{Characteristics of the cSB galaxies. 
  {\sl Column {a:}}  Behaviour of the index D$_{n}$(4000) (indicator of stellar age) along the galactic radius. The grey points denote values of the D$_{n}$(4000) index
  in the individual spaxels, the red circles mark the median values of the D$_{n}$(4000) index in bins of 0.05 in the fractional radius $R/R_{25}$, the bars show a scatter in the
  values of the D$_{n}$(4000) index about the median value in the bins, and the blue dashed line shows the median value of the D$_{n}$(4000) index in all the spaxels within the optical radius.
  {\sl Column  {b:}} Surface brightness in the emission H$\alpha$ line as a function of radius. 
  The de-reddened and corrected for the galaxy inclination surface brightness are in units of $L_{\sun}$/pc$^{2}$. The spaxels with the  H\,{\sc ii} region-like spectra are shown
  by the blue circles, the spaxels with AGN-like spectra are denoted by the dark crosses, and the spaxels with the intermediate (INT) spectra are marked by the red plus signs.
  The dashed line shows the median value of the  surface brightness in the emission H$\alpha$ line in all the spaxels within the optical radius.  
  {\sl Column  {c:}} Radial distribution of the oxygen abundance. The grey points denote oxygen abundances in the individual spaxels. The red circles mark the median
  values of the O/H in bins of 0.05 in the fractional radius $R/R_{25}$, and the bars show the scatter in the O/H  about the median value in the bins. The lines show the relations
  we determined for the radial abundance distributions. {The spaxels which show the H\,{\sc ii} region-like spectra according to the BPT classification but not locate in the
    SF area in the WHaD diagram are marked by green plus signs.}  
  {\sl Column  {d:}} Distribution of spaxels with spectra of different BPT types across the galactic image.
  The BPT radiation types for individual spaxels are colour-coded. The yellow circle shows the kinematic centre of the galaxy, the line indicates the position of the major
  kinematic axis of the galaxy, and the ellipse is the optical radius. 
}
\label{figure:app-fig1}
\end{figure*}

%===============    Fig  No App 2        
\begin{figure*}
\resizebox{0.90\hsize}{!}{\includegraphics[angle=000]{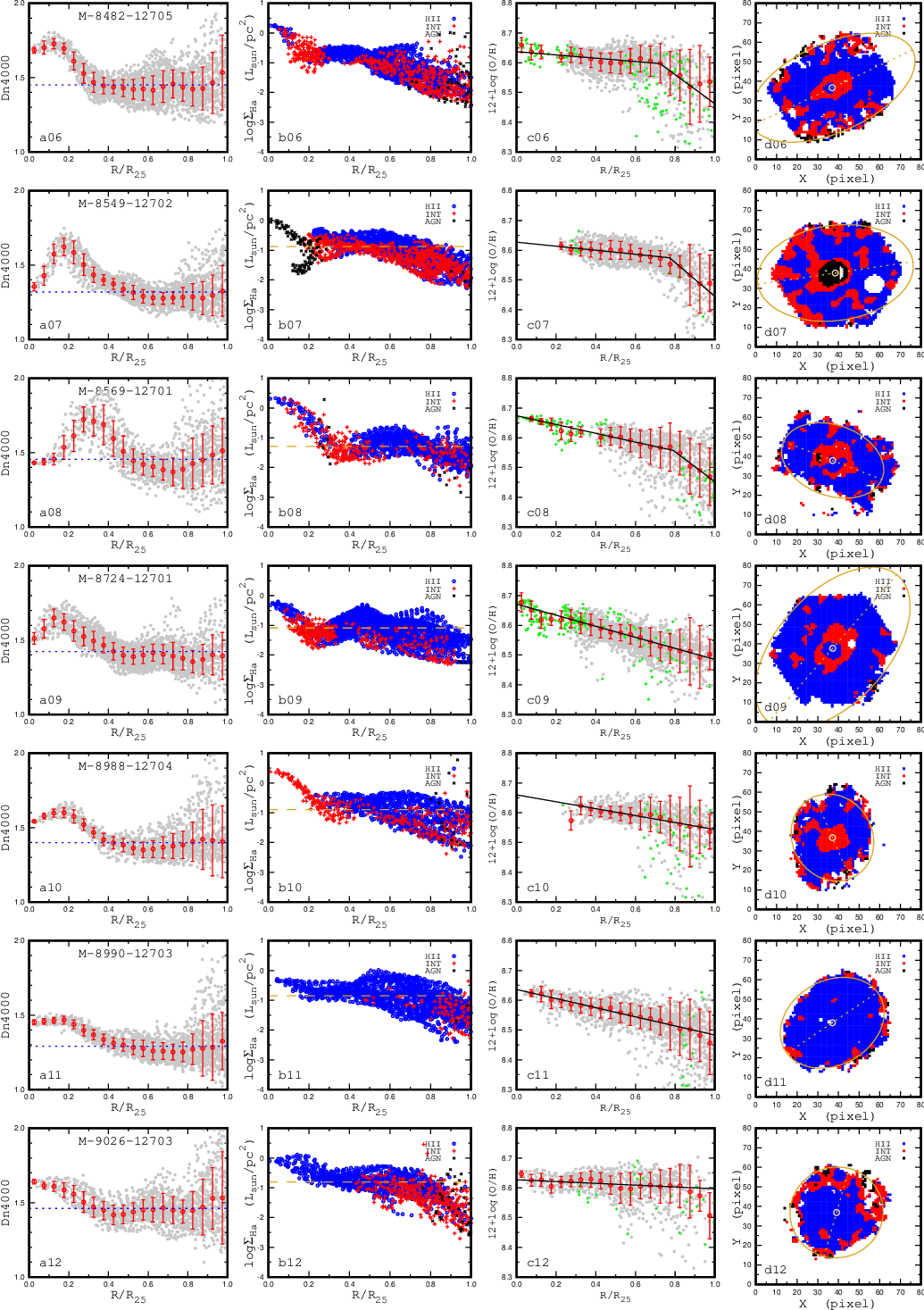}}
\caption{
  Same as Fig.~\ref{figure:app-fig1} but for other galaxies.
}  
  \label{figure:app-fig2}
\end{figure*}

%===============    Fig  No App 3        
\begin{figure*}
\resizebox{0.90\hsize}{!}{\includegraphics[angle=000]{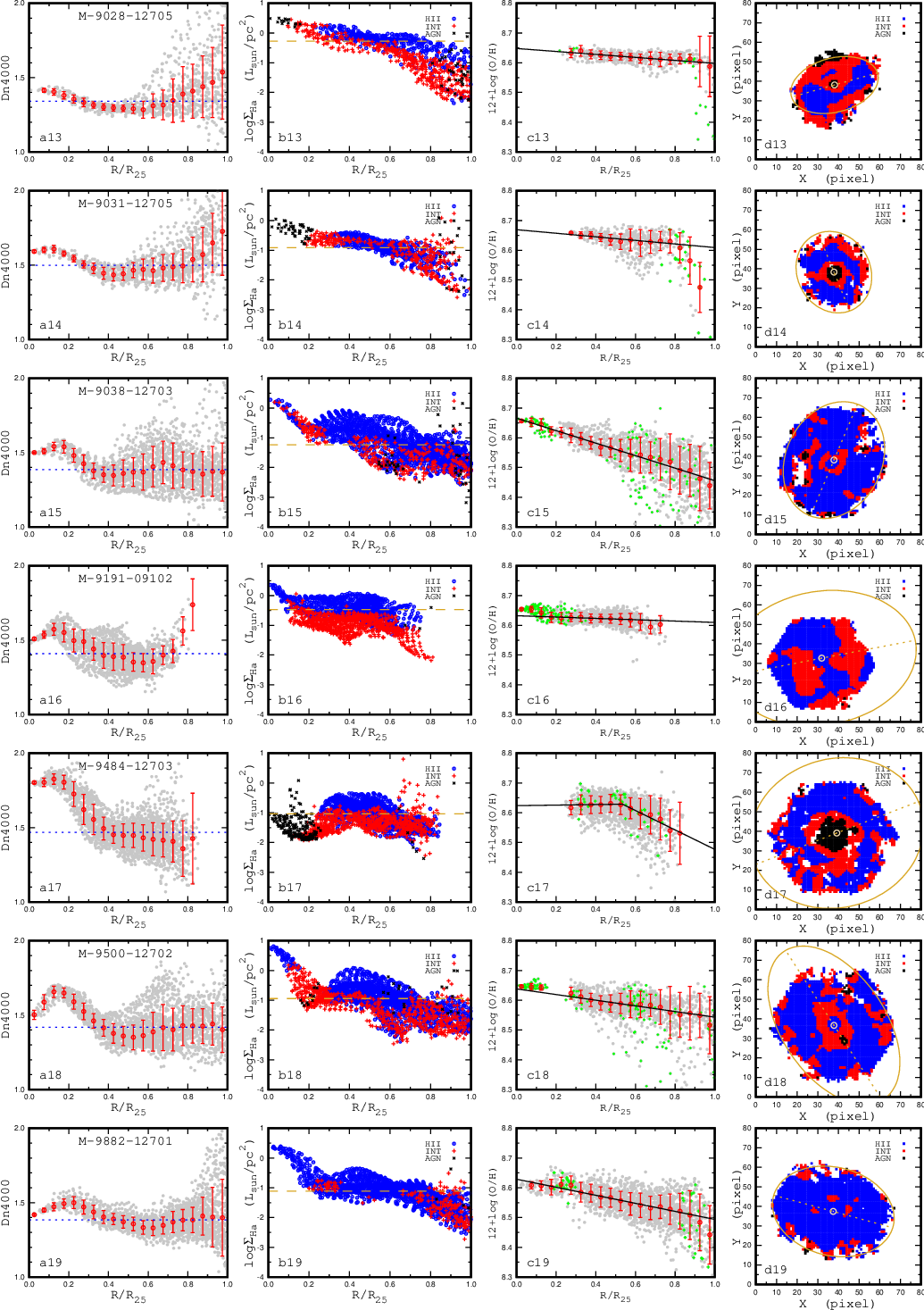}}
\caption{
  Same as Fig.~\ref{figure:app-fig1} but for other galaxies.
}  
\label{figure:app-fig3}
\end{figure*}

%===============    Fig  No App 4        
\begin{figure*}
\resizebox{0.90\hsize}{!}{\includegraphics[angle=000]{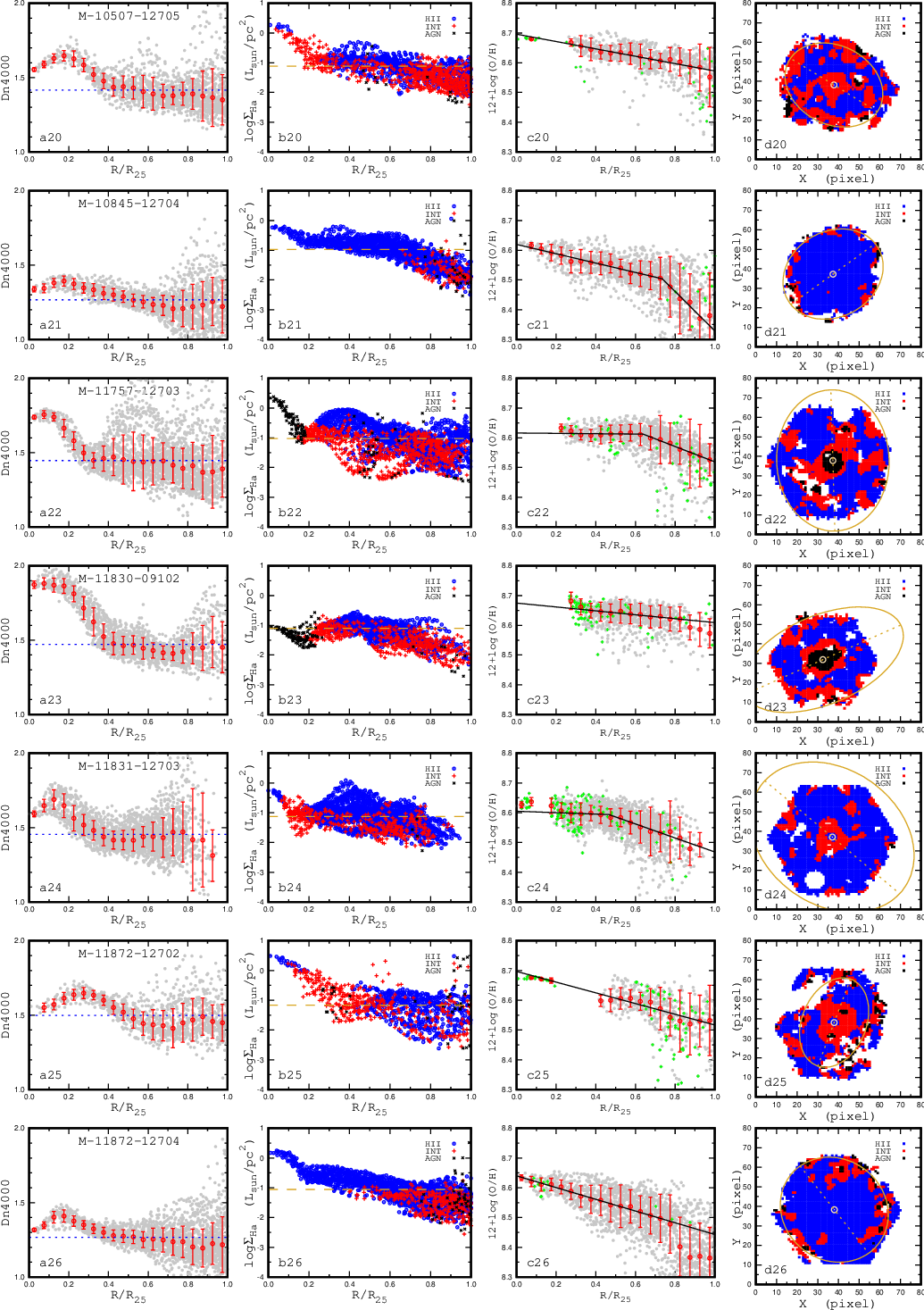}}
\caption{
  Same as Fig.~\ref{figure:app-fig1} but for other galaxies.
}  
\label{figure:app-fig4}
\end{figure*}

%===============    Fig  No App 5        
\begin{figure*}
\resizebox{0.90\hsize}{!}{\includegraphics[angle=000]{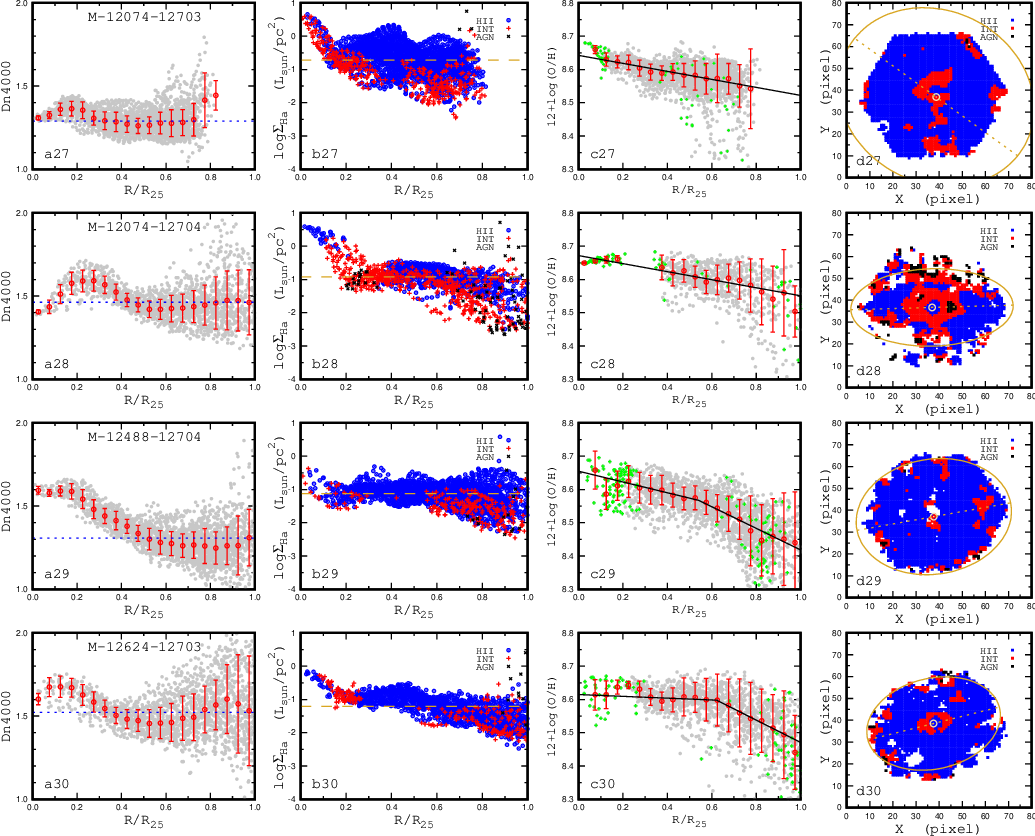}}
\caption{
  Same as Fig.~\ref{figure:app-fig1} but for other galaxies.
}  
\label{figure:app-fig5}
\end{figure*}

\clearpage

\section{Comparison between the BPT and WHaD classifications of the spaxel spectra in the cSB galaxies}
%=====================

\begin{figure*}
\resizebox{1.00\hsize}{!}{\includegraphics[angle=000]{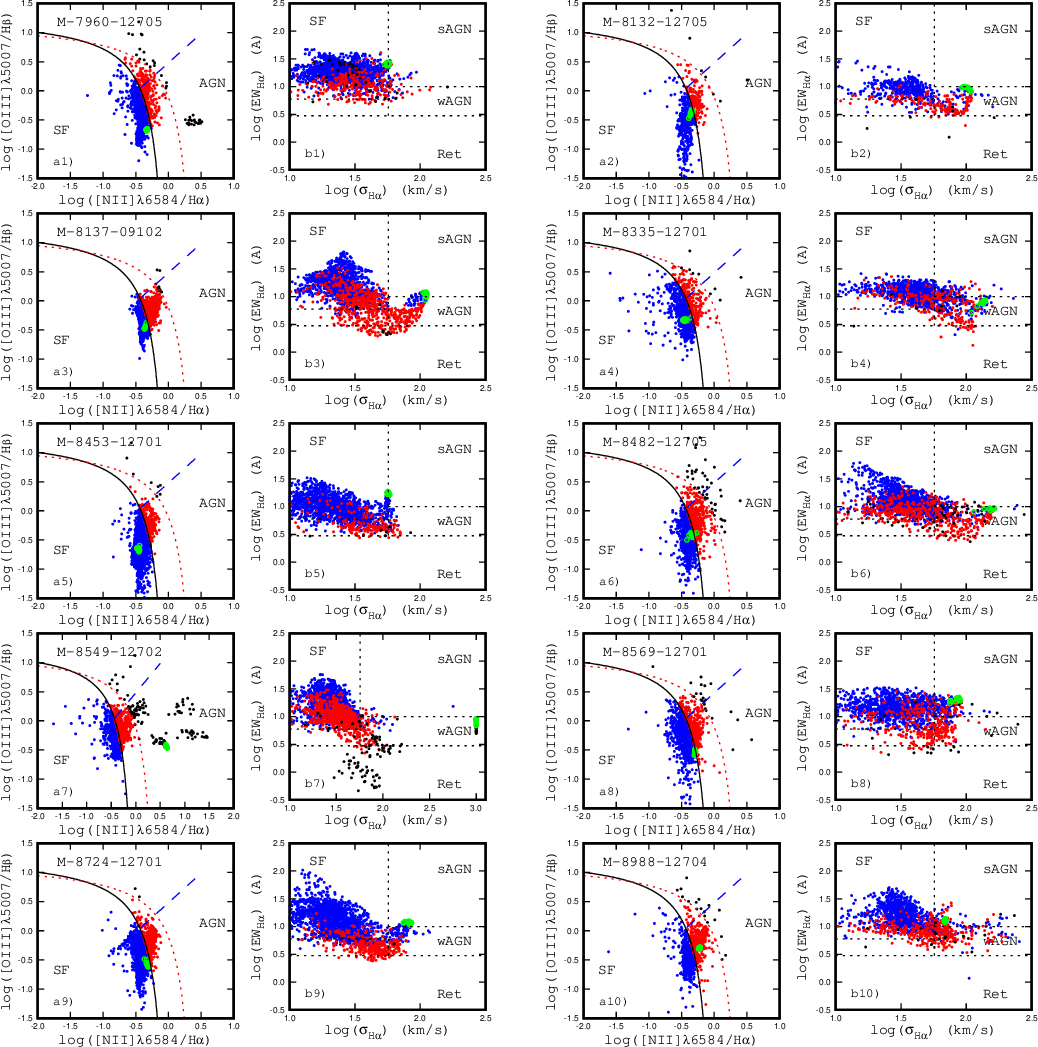}}
\caption{Two classification diagrams for the ionising sources for the individual spaxels of the cSB galaxies. Each galaxy is shown it two panels.   
  {\em Panels} {a:}  Standard BPT diagnostic diagram,  the [O\,{\sc iii}]$\lambda$5007/H$\beta$ versus \ [N\,{\sc ii}]$\lambda$6584/H$\alpha$, \citep{Baldwin1981},
    for the individual spaxels in a galaxy.
    The spaxels with the H\,{\sc ii}-region-like spectra are denoted by the blue symbols, the spaxels with AGN-like spectra are shown  by the dark symbols, and the red symbols
    are the spaxels with intermediate spectra. 
    The solid and short-dashed curves mark the demarcation line between AGNs and H\,{\sc ii}
    regions defined by \citet{Kauffmann2003} and \citet{Kewley2001}, respectively. The long-dashed line is the dividing line between Seyfert galaxies and LINERs defined by
    \citet{CidFernandes2010}.
  {\em Panels} {b:} WHaD diagram \citep{Sanchez2024} showing the equivalent-width of H$\alpha$ emission line, EW(H$\alpha$), versus the  H$\alpha$ velocity dispersion,
  $\sigma_{{\rm H}\alpha}$, for the individual spaxels in a galaxy. The colour-codes of the spaxels come from panel (a). The dotted lines divide areas of different ionising sources:
  SF, ionisation due to young-massive OB stars, related to recent star-formation activity;
  sAGNs/wAGNs, ionisation due to strong (weak) AGNs, and other sources of ionisation such as high velocity shocks;
  Ret, ionisation due to hot old low-mass evolved stars (post-AGBs), associated with retired regions within galaxies (in which there is no star-formation).
  In both panels, ten spaxels nearest to the centre of the galaxy are marked by the green circles.
}
\label{figure:appb-fig1}
\end{figure*}

\begin{figure*}
\resizebox{1.00\hsize}{!}{\includegraphics[angle=000]{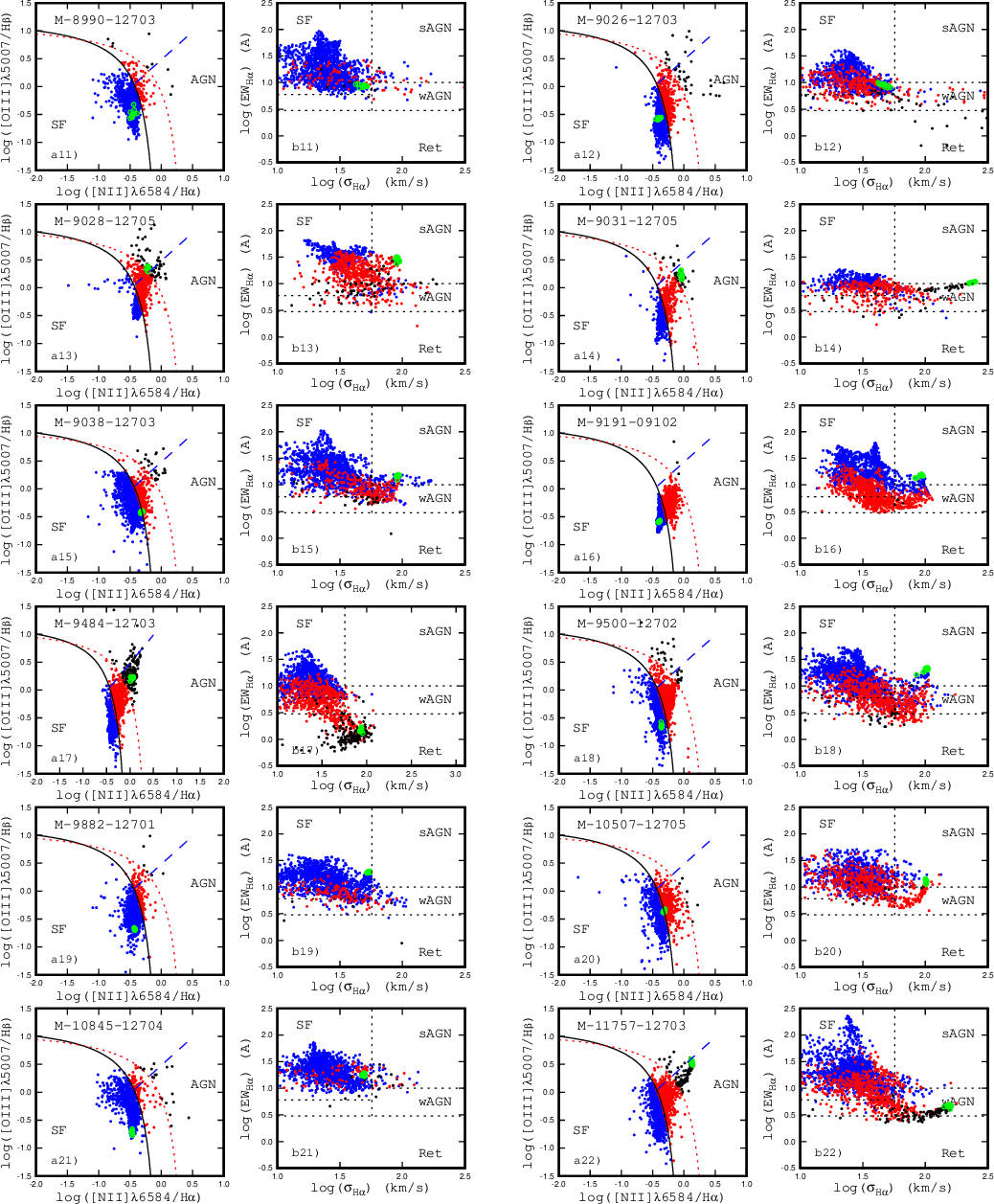}}
\caption{
  Same as Fig.~\ref{figure:appb-fig1} but for other galaxies.
}
\label{figure:appb-fig2}
\end{figure*}

\begin{figure*}
\resizebox{1.00\hsize}{!}{\includegraphics[angle=000]{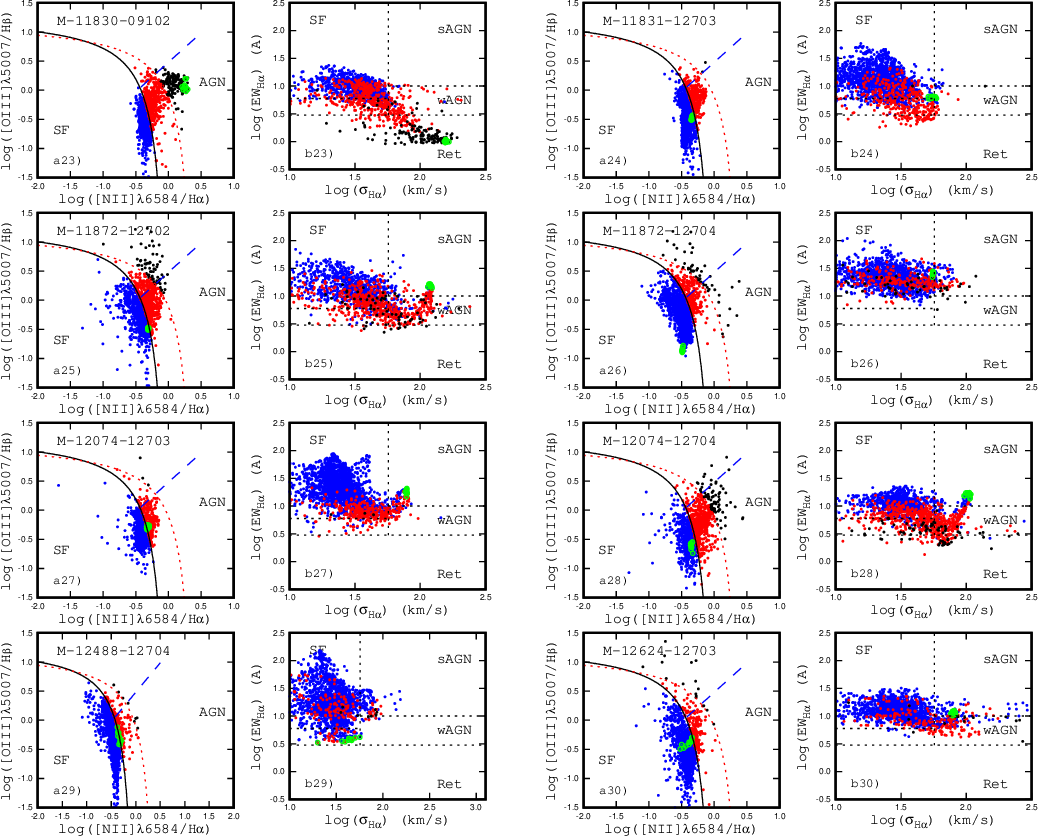}}
\caption{
  Same as Fig.~\ref{figure:appb-fig1} but for other galaxies.
}
\label{figure:appb-fig3}
\end{figure*}

Figures~\ref{figure:appb-fig1} --  \ref{figure:appb-fig3} show the comparison between the BPT and WHaD classifications of the ionising sources for the spaxel spectra
in the cSB galaxies. Each galaxy is presented in two panels. 
Panel (a) shows the BPT diagram for the individual spaxels in a galaxy. The spaxels with the H\,{\sc ii}-region-like spectra are denoted by the blue symbols,
the spaxels with AGN-like spectra are shown  by the dark symbols, and the red symbols are the spaxels with intermediate spectra. The solid and short-dashed curves mark
the demarcation line between AGNs and H\,{\sc ii} regions defined by \citet{Kauffmann2003} and \citet{Kewley2001}, respectively. The long-dashed line is the
dividing line between Seyfert galaxies and LINERs defined by \citet{CidFernandes2010}.
Panel (b) shows the  equivalent width of the emission H$\alpha$ line, $EW_{{\rm H}\alpha}$, versus gas velocity dispersion, $\sigma_{{\rm h}\alpha}$, diagram (WHaD diagram)
suggested by \citet{Sanchez2024}. They defined different areas in which the ionising source could be classified as:
(1) SF, ionisation due to young-massive OB stars, related to recent star-formation activity ($EW_{{\rm H}\alpha} >$ 6 {\AA} and $\sigma_{{\rm h}\alpha} <$ 57 km/s) 
(2) sAGNs/wAGNs, ionisation due to strong (weak) AGNs, and other sources of ionisations like high velocity shocks ($EW_{{\rm H}\alpha} >$ 10 {\AA} and $\sigma_{{\rm h}\alpha} >$ 57 km/s
for sAGN and  (10 {\AA} $> EW_{{\rm H}\alpha} >$ 3 {\AA} and $\sigma_{{\rm h}\alpha} >$ 57 km/s for wAGN); and (3) Ret, ionisation due to hot old low-mass evolved stars (post-AGBs), associated
with retired regions within galaxies, in which there is no star-formation, ($EW_{{\rm H}\alpha} <$ 3 {\AA}).
The colour of each spaxel corresponds to the BPT classification (comes from panel (a)).
In both panels, ten spaxels nearest to the centre of the galaxy are marked by the green circles.

%===============    Fig  No           BPT and WHaD classification diagrams 

%===============    Fig  No           BPT and WHaD classification diagrams 

%===============    Fig  No           BPT and WHaD classification diagrams 

%\end{appendix}
%=========

\end{document}